\documentclass[12pt]{iopart} %for iopart + ams packages
\expandafter\let\csname equation*\endcsname\relax
\expandafter\let\csname endequation*\endcsname\relax
\usepackage{amsmath,amsthm,amssymb,amsfonts,mathrsfs, mathtools}
\usepackage{array}
\usepackage{graphics}
\usepackage{aas_macros}
\usepackage{graphicx}
\usepackage{hyperref}
\usepackage[inline]{enumitem}
\usepackage{wasysym} % for astronomy symbols
\usepackage{bm} % for bold greek symbols
\usepackage{cancel}
\usepackage{float}
\usepackage{caption}
\usepackage{subcaption}

\renewcommand{\rmd}{\ensuremath{\mathrm{d}}}
\renewcommand{\rme}{\ensuremath{\mathrm{e}}}
\renewcommand{\vec}[1]{\ensuremath{\bm{#1}}}
\newcommand{\np}{\ensuremath{\mathrm{NP}}}

\newcommand{\f}[2]{\ensuremath{\frac{#1}{#2}} }
\newcommand{\deriv}[2]{\ensuremath{\frac{\rmd #1 }{\rmd #2}}}
\newcommand{\sderiv}[2]{\ensuremath{\frac{\rmd^2 #1 }{\rmd #2^2 }}}

\newcommand{\myfigref}[2]{Figure~\ref{#1}(\subref{#2})}

\begin{document}

\title{The Geometrical Structure of the Tolman~VII solution}
\author{Ambrish M. Raghoonundun} 
\author{David W. Hobill}

\begin{abstract}
  The Tolman~VII solution, an exact analytic solution to the
  spherically symmetric, static Einstein equations with a perfect
  fluid source, has many characteristics that make it interesting for
  modelling high density physical astronomical objects.  Here we
  supplement those characteristics with the geometrical tensors that
  this solution possess, and find that the Weyl, Riemann, and Ricci
  tensor components show unexpected mathematical behaviour that change
  depending on physically motivated parameters, even though the mass
  of the modelled objects is fixed.  We show these features firstly
  through tensor components, and then through the scalars in the null
  tetrad formalism of Newmann and Penrose.  The salient conclusion of
  this analysis is the intimate relationship between the Tolman~VII
  solution and the constant density Schwarzschild interior solution:
  the former being a straight forward generalization of the latter
  while eschewing the unphysical constant density.
\end{abstract}

\maketitle
%\ioptwocol 
\section{\label{sec:intro}Introduction}
The Tolman--Oppenheimer--Volkoff (TOV) equation is commonly used to
construct static spherically symmetric solutions of self-gravitating
objects consistent with general relativity.  The TOV equation is
particularly useful for studies of white dwarfs, neutron stars, and
quark stars.  This equation can be seen as a generalization of the
Lane-Emden equation of Newtonian stars to general relativistic ones.
If the aim of the study is the prediction of masses and radii of
compact objects this is where the analysis usually terminates, since
this type of analysis is used to constrain equations of
state (EOS) of matter in the high density regimes.  The usual view is
that since the TOV equation is derived from general relativistic
considerations, a proper general relativistic model incorporating all
of known physics has been found.  In this article the goal is to show that
even if one is given an EOS, a host of other geometrical properties
that are not readily apparent can also be obtained.  What will allow
such an analysis is the availability of an exact closed form solution
that also yields a physical EOS.  By inverting the flow of the
historical derivation of this solution, one can show how starting with
an EOS that is well behaved, geometrical tensors can be generated.
The geometrical information presented here will be in the Riemann and
Weyl tensors and in the Ricci tensor and scalar.  Together with the
metric these provide a complete geometrical description of the
behaviour of the solution, and since the solution being discussed is
physical~\cite{Rag09, RagHob15}, the description is hopefully
applicable to the real world.

The point of view to be pursued in this article is that the exterior
vacuum geometry is described by the Schwarzschild exterior metric due
to some central object whose interior consists of a perfect fluid.
The exterior field is determined solely by the Schwarzschild mass seen
outside, but what will be shown here is that even with a unique
interior solution modulo parameters, the behaviour, and qualities of
the material inside can be very different. The geometrical tensors are
found to be parameter dependent, and indeed one expects the Ricci
tensor \(R_{ab}\)
components, and the Ricci scalar \(R,\)
to change with the interior solution parameters. That the Weyl tensor
\(C_{abcd}\)
components which usually encode the free gravitational field has a
different behaviour depending on these parameters is perhaps
surprising.

This article is organized as follows: following a brief description of
the solution in section~\ref{sec:tvii}, a brief review of the
different definitions of mass in general relativity is given in
section~\ref{sec:mass}.  The fixing of certain parameters is thus
achieved through these definitions before continuing with a thorough
geometrical analysis of the solution.  The behaviour of the metric and
other tensors is discussed in section~\ref{sec:mgstruct}.  Then two
different tetrads are introduced: an orthogonal one in
section~\ref{ssec:tetradOrtho} and a null one in
section~\ref{ssec:tetradNull}, to express the tensor quantities found
previously in the different tetrad bases, before discussing the
results in section~\ref{sec:discussion}.  Finally a set of expressions
and graphs of the tetrad formalism quantities is provided in
the appendix

\section{\label{sec:tvii}The Tolman VII solution}
The Tolman~VII solution was generated through a mathematical ansatz on
the static, spherically symmetric Einstein's equations by
Tolman~\cite{Tol39} in 1939.  Several authors~\cite{DelLak98,
  FinSke98} subsequently proved that it was a viable \emph{physical}
solution, and showed~\cite{RagHob15} that it could be used
advantageously to model compact astrophysical objects.  An interesting
aspect of the Tolman~VII solution is that an exact analytic equation
of state can be obtained for it.  Since a closed form solution exists,
all geometrical quantities of interest can be computed directly.  The
behaviour of these we claim gives insight into exactly which parts of
the external field is influenced by which internal contributions.  In
particular it is found that the Weyl tensor contribution in the
interior is only sometimes continuous with the pure Weyl Schwarzschild
exterior, and that the Ricci tensor components which is always zero
outside the star, has interesting non-monotonic behaviour inside the
star, pointing towards non-intuitive curvature behaviour in the
interior of stars.

The Tolman~VII solution is completely specified by the two metric
functions \(Z(r) = \rme^{-\lambda(r)}\)
and \(Y(r) = \rme^{\nu(r)/2}\)
of the spherically symmetric and static line element
\begin{equation*}
  \label{eq:lineelement}
  \rmd s^{2} = \rme^{\nu(r)} \rmd t^{2} - \rme^{\lambda(r)} \rmd r^{2} - r^{2} \rmd \theta^{2} -r^{2} \sin^{2}\theta \rmd \varphi^{2},
\end{equation*}
a mass density function \(\rho\), and an isotropic pressure \(p:\)
all of which are functions of \(r,\) the Schwarzschild radial coordinate.

A \((+,-,-,-)\) signature is used along with geometrical units where \(G=c=1.\)
Commencing with a quadratic mass density function of the form
\begin{equation}
\label{eq:Density}
\rho(r) = \rho_{c}\left[ 1 - \mu \left(
    \f{r}{r_{b}}\right)^{2}\right],
\end{equation}
the three constants, \(\rho_{c}\)
being the central density at \(r=0\)
in the star, while \(r_{b}\)
is the coordinate radius of the boundary, and \(\mu\)
is a ``self-boundedness'' parameter that allows for the surface density
to be zero when \(\mu=1\)
and non-zero for \( 0 \leq \mu < 1,\)
as will become clear in what follows; the Einstein equations lead
directly to the first metric function,
\begin{equation}
  \label{eq:Z}
Z(r) = 1 - \left( \f{\kappa \rho_{c}}{3}\right) r^{2} + 
  \left(\f{\kappa \mu \rho_{c}}{5r_{b}^{2}}\right) r^{4} \eqqcolon 1- br^2+ ar^4.
\end{equation}
With a little bit more effort (see for
example~\cite{Iva02,RagHob15}) the second metric function given by
\begin{equation}
  \label{eq:Y}
 Y(r)  = c_{1} \cos (\phi \xi(r)) + c_{2} \sin (\phi \xi(r)) , \quad \text{with } 
    \phi = \sqrt{\f{a}{4}}.
\end{equation}
can be obtained, from which simple manipulations of the Einstein equations lead to 
the very complicated looking expression for the pressure:  
\begin{equation}
  \label{eq:PressureR}
  \kappa p = \f{4\phi [c_{2} \cos{(\phi\xi)} - c_{1} \sin{(\phi\xi)}] \sqrt{1 - br^{2} + ar^{4}}}{c_{1}\cos{(\phi\xi)} + c_{2} \sin{(\phi\xi)}} - 4ar^{2} + 2b - \kappa \rho,
\end{equation}
where the transformed radial coordinate \(\xi\) whose expression is
\begin{equation}
  \label{eq:xiCoth}
  \xi(r) = \f{2}{\sqrt{a}} \coth^{-1} \left( \f{1+\sqrt{1-br^2+ar^4}}{r^2\sqrt{a}}\right),
\end{equation}
has been used.
The complicated form of this pressure is the reason Tolman initially
abandoned~\cite{Tol39} this solution, but it was shown
previously~\cite{RagHob15,DurRaw79,DurGeh71} that this function has
some very rich physical properties.

The constants \(c_{1}\) and \(c_{2}\) are fully determined through
\begin{subequations}
\begin{align}
  \label{eq:c1,c2}
  c_{1} &= \gamma \cos{(\phi\xi_{b})} - \f{\alpha}{\phi} \sin{(\phi \xi_{b})}, \\
  c_{2} &= \gamma \sin{(\phi\xi_{b})} + \f{\alpha}{\phi} \cos{(\phi \xi_{b})},
\end{align}  
\end{subequations}
once boundary conditions are imposed, where
\begin{align*}
  \alpha &= \f{\kappa \rho_{c}}{4}\left( \f{1}{3} - \f{\mu}{5}\right), \\
  \gamma &= \sqrt{1 - \kappa\rho_{c}r^{2}_{b} \left( \f{1}{3} - \f{\mu}{5}\right)},
\end{align*}
while \(a,\)
and \(b\)
are constants that are defined through equation~\eqref{eq:Z}.  As for
\(\xi_{b},\)
it is a shorthand for the value of the function \(\xi(r)\)
which was provided in equation~\eqref{eq:xiCoth}, evaluated at the
boundary, \(r=r_{b}.\)

At this juncture, it will be beneficial to note that this form for the
\(Z\)
metric function can be seen as a generalization of the Schwarzschild
interior solution: indeed, were one to only use the first two terms in
the expression of \(Z,\)
the end result would be the exact form of the Schwarzschild interior
solution's \(Z\)
function: this parallel between Tolman~VII and Schwarzschild interior
will become clearer in this article.  This generalization relation has
been erroneously interpreted previously in~\cite{NeaIshLak01}, where
it was stated that the limit of the equivalent of \(\mu=0\)
does not yield the Schwarzschild interior solution.  The is untrue,
but the issue relates to how the differential equation for \(Y\)
is solved.  The final form of the equation to be solved is a harmonic
oscillator type equation whose solutions for zero frequency
corresponds to the Schwarzschild interior solution when \(\mu = 0.\)
The Tolman~VII solution by contrast corresponds to positive frequency
trigonometric solutions whose limit when \(\mu \to 0,\)
does not give the linear solution as the frequency goes to zero.  The
similarity to the Schwarzschild interior solution is also evident in
the equation for the density~\eqref{eq:Density} in the limit of
\(\mu=0\)
when one gets the constant density solution -- a well know
characteristic of Schwarzschild interior.

Since the vacuum region is spherically symmetric and static, the only
candidate by Birkhoff's theorem is the Schwarzschild exterior
solution.  The Israel-Darmois junction conditions for this system can
then be shown to be equivalent to the following two
conditions~\cite{Syn60}:\begin{subequations}
  \label{eq:Boundary1+2}
  \begin{align}
    \label{eq:BoundaryP}
    p(r_{b}) &= 0, \quad \text{and,} \\
    \label{eq:BoundaryZ}
    Z(r_{b}) &= 1-\f{2M}{r_{b}} = Y^{2}(r_{b}).
  \end{align}
\end{subequations}
Where \(M = m(r_{b})\) is the total mass of the sphere as seen by an outside
observer, and \(m(r)\) is the mass function defined by
\begin{equation}
  \label{eq:MassFunction}
  m(r) = 4\pi \int_{0}^{r} \rho(\bar{r}) \bar{r}^{2 }\rmd \bar{r},
\end{equation}
and \(p(r),\)
and \(\rho(r)\)
are respectively the pressure and density of the perfect fluid used to
model the fluid interior.  Furthermore requiring the regularity of the
mass function, that is the mass function vanishes at the \(r=0\)
coordinate from physical considerations: \(m(r=0) = 0,\)
fixes the value of the integration constants uniquely.

\begin{figure}[h!]
\includegraphics[width=\linewidth]{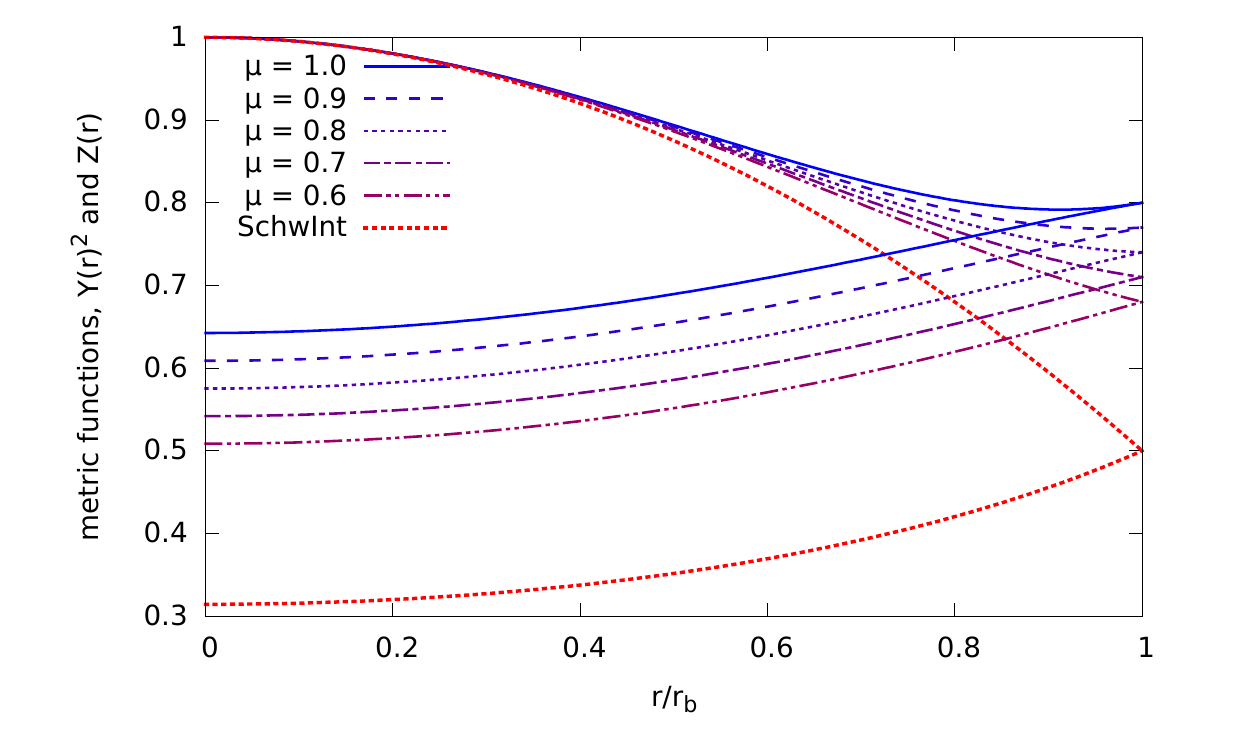}
\caption{(Colour online) The $Z$ and $Y^{2}$ metric functions with the
  radial coordinate inside the star for the TVII solution.  The $Z(r)$
  functions can be identified by the fact that $Z(r=0)=1,$ for all
  $\mu.$ The parameter values are $\rho_{c}=3/(16\pi)$ and
  $\mu$ taking the various values shown in the legend.  Also shown for
  comparison is the Schwarzschild interior solution's metric functions
  abiding by the same boundary conditions.  Visually the trend of
  $\lim_{\mu \to 0}(\mathrm{TVII}) \to \mathrm{SchwInt}$ is clear.}
\label{fig:metricmatching}
\end{figure}

\begin{figure}[h!]
\centering
   \begin{subfigure}[b]{0.5\textwidth}
       \includegraphics[width=\textwidth]{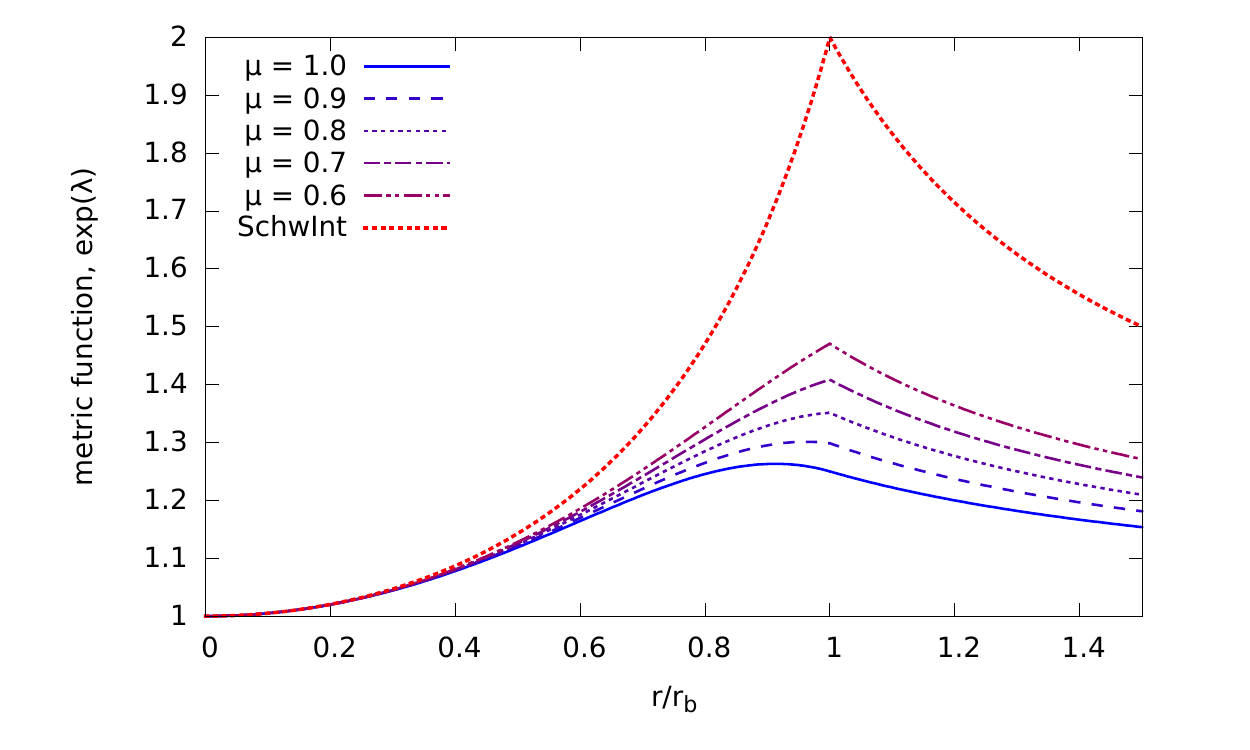}
       \caption{The $\lambda$ metric function}\label{fig:lambda}
   \end{subfigure}~
   \begin{subfigure}[b]{0.5\textwidth}
       \includegraphics[width=\textwidth]{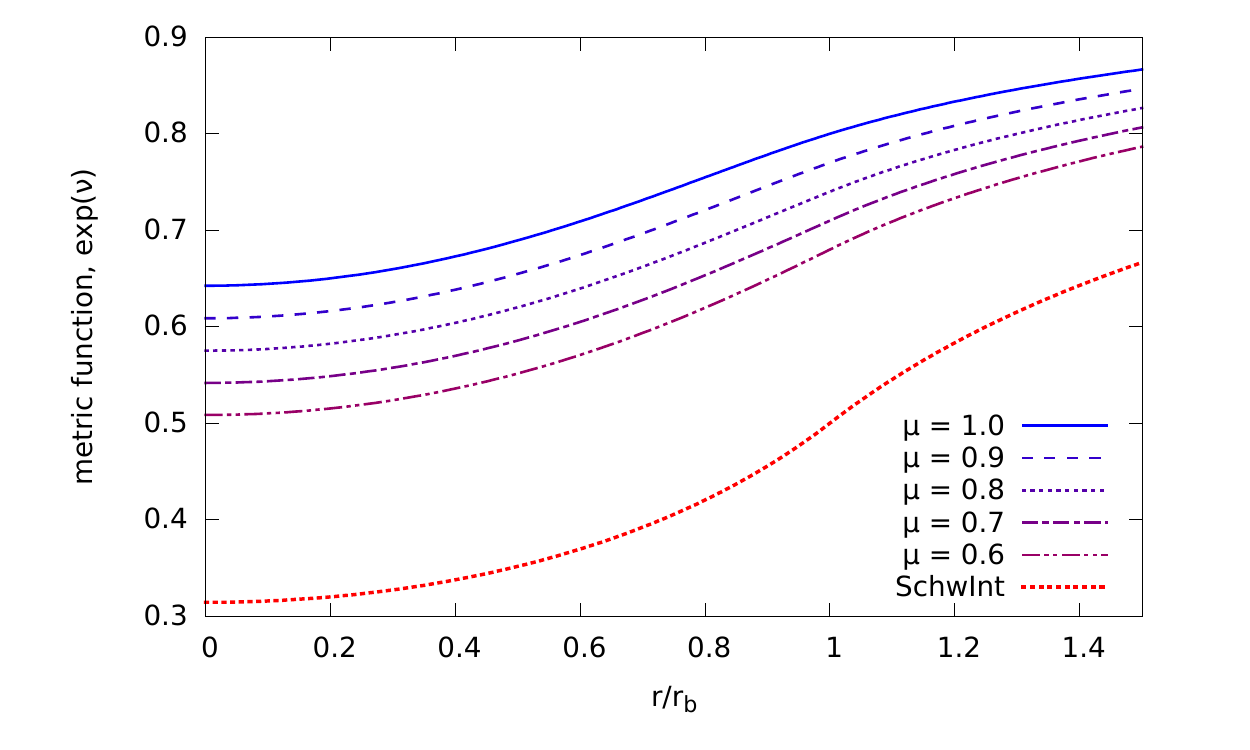}
       \caption{The $\nu$ metric function}\label{fig:nu}
   \end{subfigure} 
\centering
\caption{(Colour online) The metric functions with the radial
  coordinate inside and outside the star for the TVII solution. The
  parameter values are $\rho_{c}=3/(16\pi)$ and $\mu$ taking
  the various values shown in the legend.  The Schwarzschild interior
  metric for the same parameter values are also shown for
  comparison. Again the unmistakable trend of $\lim_{\mu \to 0}$ of
  TVII resulting in Schwarzschild interior is clear. The masses
  associated with these solutions are all different.}
\label{fig:lambdaNu}
\end{figure}

Enforcing the boundary conditions ensures the matching of the value of
the metric coefficients at the boundary as is shown in
Figure~\ref{fig:metricmatching},
leading to matching to a Schwarzschild exterior metric which is also
shown next in~\myfigref{fig:lambdaNu}{fig:lambda} and
in~\myfigref{fig:lambdaNu}{fig:nu}.  In all the figures the
Schwarzschild Interior solution, which was generated with the same
parameter values as the Tolman~VII solution is also shown, in an
attempt to show the similarities between them.  Now that the boundary
conditions have been enforced, and hence a full solution to the
Einstein equations been obtained, one can begin to investigate the
geometrical structure of this solution.  However this endeavour is
made difficult by the presence of \emph{three} different parameters
that can be changed simultaneously: \(\rho_{c}, r_{b},\)
and \(\mu.\)
Each of these change the character of the interior solution and the
EOS quite drastically.  This effect on the latter was shown previously
in~\cite{RagHob15}, whereas in this manuscript, being mainly interested in
the former aspects, we proceed by fixing the only parameters usually
available (measured) for these astrophysical objects: the mass and the
radius.

\section{\label{sec:mass} Mass in general relativity}
In what follows, the Tolman VII solution is compared and contrasted to
the Schwarzschild interior solution.  In order to simplify the
analysis, certain parameters will be fixed while others will be
allowed to vary.  The parameter that was kept fixed in the following
is the external mass perceived at infinity: the mass that is used in
the externally matched Schwarzschild exterior solution: typically
\(M=1/4\)
was used in all the figures.  Through the use of geometric units, the
higher Buchdahl~\cite{Buc59} limit of \(M=4/9,\)
can be used directly(without unit conversions) and thus the use of
\(M=1/4\)
throughout gives some leeway in changing other parameters, while
skirting around possible singular behaviour in the metric function.

This choice does not fix the solution uniquely, and in what follows
either \begin{enumerate*} \item the radius is fixed to some specific
  normalized value: typically to \(r_{b} = 1 ,\)
  and in so doing one varies the central density\(\rho_{c}\)
  to achieve the same external mass, or \item the central density is
  fixed to \(\rho_{c} = 3/(16\pi)\)
  and the radius \(r_{b}\)
  of the sphere varied, while still keeping the mass fixed.  The one
  radius that is fixed for comparison purposes is the outer boundary
  of the Schwarzschild interior solution which is set to unity, in this
  case too.
\end{enumerate*}
In either case, the dimensionless parameter \(\mu,\)
is varied to give comparisons to the Schwarzschild interior
solution.  As a result the following trends are expected to be visible:
\begin{enumerate*}
\item since the mass is fixed, and the central density is changing while
  the radius is fixed, all discontinuities, if they exist, occur
  at \(r_{b}=1\)
  in the first set of figures~{\bf (a)}.  Decreasing \(\mu\)
  from one to zero then results in the central density decreasing so
  that one still gets a fixed mass while the boundary density can
  increase at \(r_{b}=1.\)
\item In the second set of figures~{\bf (b)}, decreasing \(\mu\)
  from one to zero, increases the boundary radius \(r_{b}\)
  as is evident by the increasing \(r\)--coordinate of the discontinuities
  of the corresponding curves shown.
\end{enumerate*} 
We now give a brief description of the different masses appearing in
the literature, to be able to distinguish between the Schwarzschild
mass \(M = m(r_{b}),\)
the Tolman--Whittaker mass \(M_{T}\)
and the proper mass \(M_{P}.\)
The mass given in equation~\eqref{eq:MassFunction} is known as the
Schwarzschild mass.  This mass depends on the density \(\rho\)
of the material source and does not take into account the
gravitational energy.  The usual method of including gravitational
energy is through what is known as the
Tolman--Whittaker(TW)--mass~\cite{Tol66, LanLif80} given by
\begin{equation}
  \label{eq:TWmass}
  M_{T} \coloneqq \int_{V} \left( T_{0}{}^{0} - T_{1}{}^{1} - T_{2}{}^{2} - T_{3}{}^{3} \right) \sqrt{-g} \rmd V 
  = 4\pi \int_{0}^{r_{b}} \left( \rho + 3p \right) \f{r^{2}Y}{\sqrt{Z}} \rmd r,
\end{equation}
where \(T_{a}{}^{b}\)
are the components of the energy momentum tensor, and \(g\)
is the metric determinant.  The method of finding this expression
involves the use of the Tolman pseudo-tensor density of gravitational
energy \(t_{a}^{b},\)
is quite lengthy, and will not be given in full here, instead the
reader is referred to~\cite[pg. 224]{Tol66} where full derivations of
this quantity is provided.

\begin{figure}[h!]
\includegraphics[width=\linewidth]{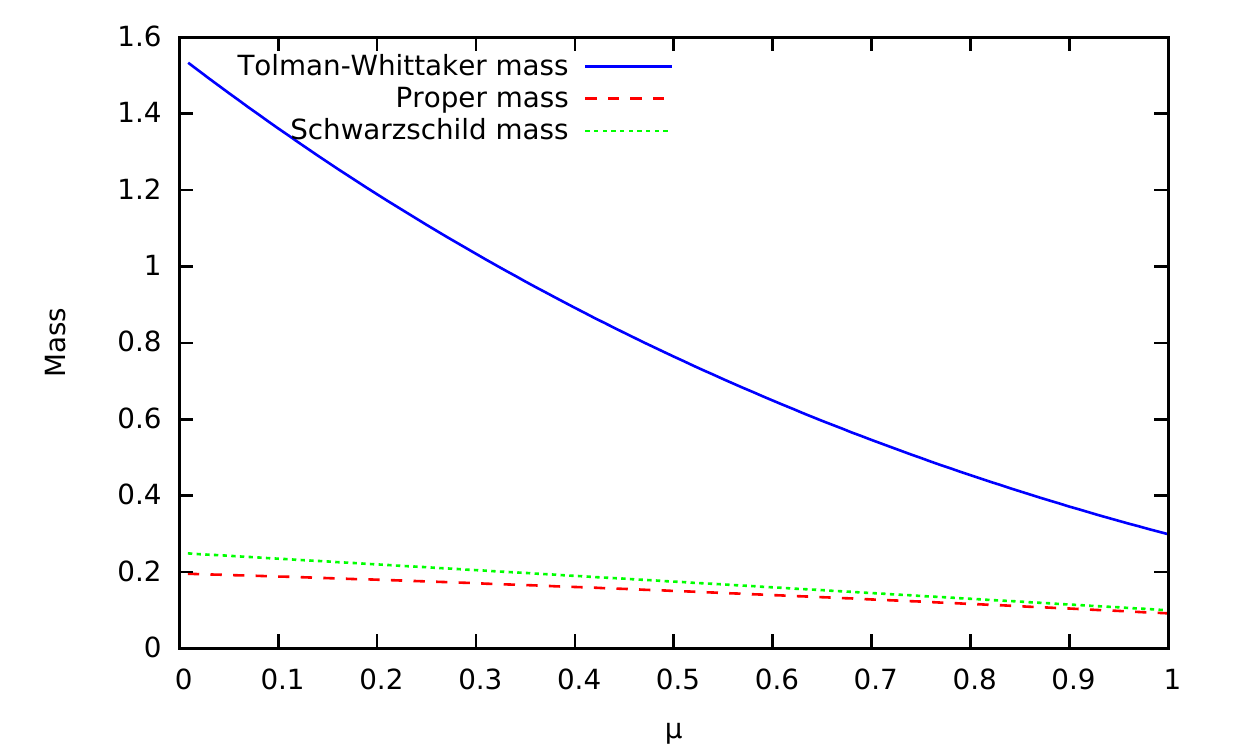}
\caption{(Colour online) The 3 different masses with changing value of
  $\mu$.  Of importance is the much higher value of the Tolman--Whittaker
  mass, since it includes all the gravitational energy of the system.
  The difference between the Schwarzschild and proper mass is not very
  large, and is due to the appearance of the metric determinant in the
  latter integral.}
\label{fig:Masses}
\end{figure}

This quantity is different from the proper mass \(M_{P},\)
which is also an integral over the proper space
volume~\cite{HaePotYak07}, but only on the energy density
\(T_{0}{}^{0},\)
that is, 
\begin{equation}
  \label{eq:DefnMP}
M_{P} \coloneqq \int_{V} \left( T_{0}{}^{0} \right) \sqrt{-g} \rmd V = \int_{V} \rho(r) \sqrt{-g} \rmd V,
\end{equation}
and is commonly used in the TOV methods' point of view where
it is compared to the total baryon rest mass. The difference between
these latter two masses yields the ``gravitational energy'' of the
star.  However this terminology will not be used in this article since
the main interest here is in the geometry of the solutions.  Both the
Schwarzschild and Tolman-Whittaker masses for these models are
calculated, and the former mass is fixed to a value of \(M=1/4\)
when appropriate, however just to provide an idea of the relative
magnitudes of the three different masses and the effect of the
parameter \(\mu\) on them, Figure~\ref{fig:Masses} is given.

The reason the Schwarzschild mass is fixed instead of the TW-mass is
practical: it is easier to compute.  Furthermore, the Buchdahl limit
is usually expressed very succinctly in terms of the Schwarzschild
mass which is also the mass that is measured at infinity by observers:
and hence one can use observed values of masses of neutron stars as a
guide in setting right magnitudes more easily.

\section{\label{sec:mgstruct}The metric and geometrical structure}
To proceed, first everything is re-expressed in terms of the mass as
seen from an external point of view, since this is the only observable
usually available to astronomers.  By fixing this external mass, the
external Schwarzschild metric is also fixed, and thus all the
different coloured lines on the right of Figure~\ref{fig:lambda} and
Figure~\ref{fig:nu} will collapse into one single line, corresponding
to the specified mass.  The internal geometrical tensors however will
be very different, while still giving the same external field.  This
rich structure will be made clear in the graphs presented, however
first one re-expresses two of the three parameters, \(\rho_{c},\)
in terms of the mass through:
\begin{equation}
  \label{eq:M.rho_c}
  \rho_{c} = \f{15M}{4\pi r_{b}^{3}(5-3\mu)},
\end{equation}
and similarly \(r_{b}\):
\begin{equation}
  \label{eq:M.r_b}
r_{b} = \sqrt[3]{\f{15M}{4\pi\rho_{c}(5-3\mu)}},
\end{equation}
where \(M\)
is the Schwarzschild mass being fixed.  Equations~\eqref{eq:M.rho_c} and~\eqref{eq:M.r_b} will allow the
investigation of the geometrical tensors inside the sphere whilst
keeping the mass constant, and still allow the flexibility of
spanning the whole solution space from ``natural'' Tolman~VII where
\(\mu=1\)
through ``self-bound'' Tolman~VII when \( 0 < \mu < 1,\)
to finally Schwarzschild interior when \(\mu=0.\)

In most of the plots, the boundary radius has been normalized at
\(r_{b} = 1\)
for the Schwarzschild interior solution.  When the boundary \(r_{b}\)
of Tolman~VII is actually varied, the different \(\mu\)
values together with the constant \(\rho_{c} = 3/(16\pi)\)
will achieve the same mass in the solutions.  However when \(r_{b}\)
is kept fixed at \(r_{b} = 1,\)
the same exterior mass is obtained through changing \(\rho_{c}.\)
In this article the causality condition which is an additional
constraint that can be imposed on the solution after the functional
form has been obtained will not be investigated.  As a result some
chosen parameter values might result in causality violations in the
solution, however causality can be imposed if it were so desired. (See
e.g.\ \cite{RagHob15} for restrictions imposed by causality).

\begin{figure}[h!]
\centering
   \begin{subfigure}[b]{0.5\textwidth}
       \includegraphics[width=\textwidth]{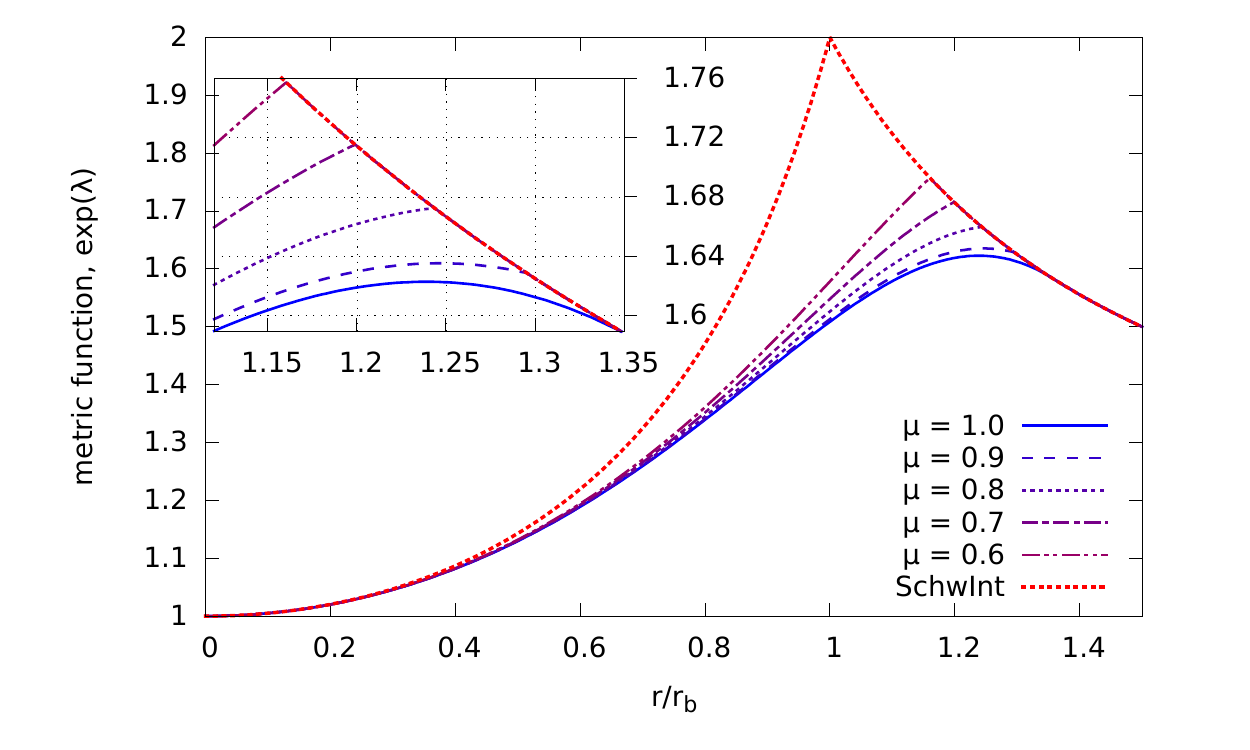}
       \caption{The $\lambda$ metric function}\label{fig:lambdaR}
   \end{subfigure}~
   \begin{subfigure}[b]{0.5\textwidth}
       \includegraphics[width=\textwidth]{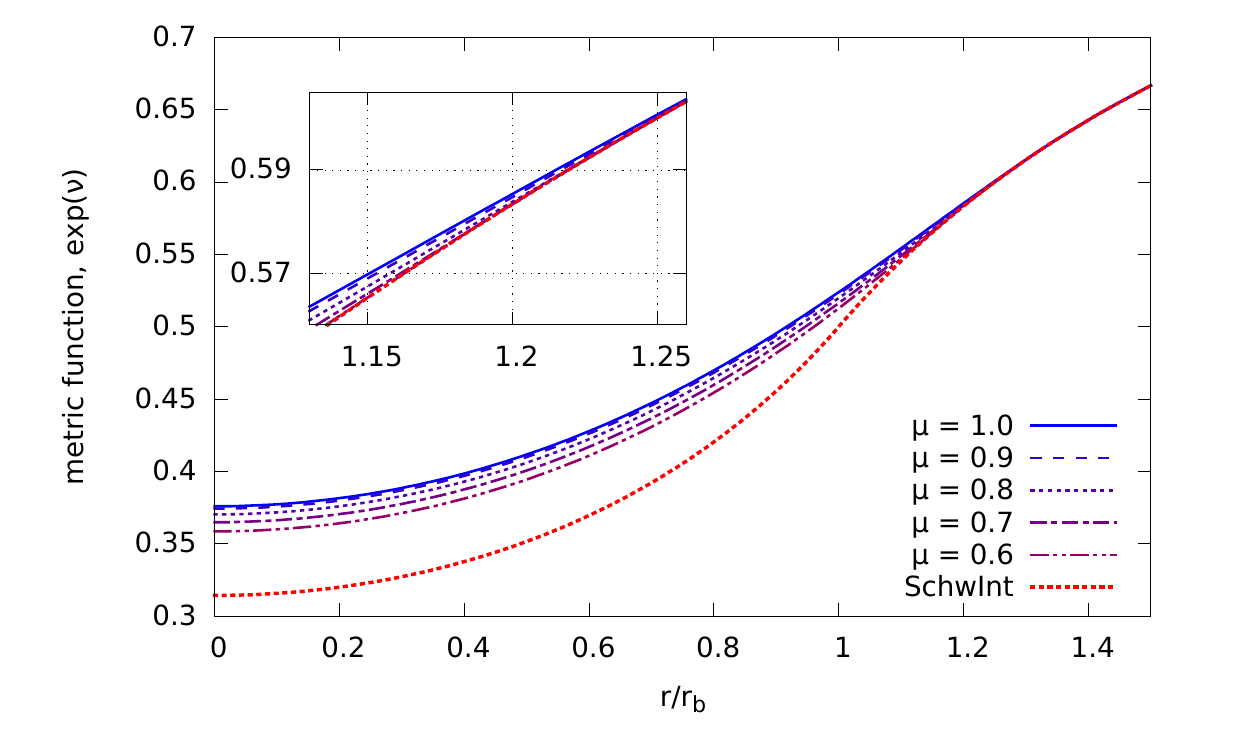}
       \caption{The $\nu$ metric function}\label{fig:nuR}
   \end{subfigure} 
\centering
\caption{(Colour online) The metric functions with the radial
  coordinate inside and outside the star for the TVII solution. The
  parameter values are $M = 1/4, \rho_{c} = 3/(16\pi)$ and $\mu$ taking
  the various values shown in the legend.  Also shown for
  comparison, the Schwarzschild interior metric for the same parameter
  values. Again the unmistakable trend of $\lim_{\mu \to 0}$ of TVII
  resulting in Schwarzschild interior is clear. The boundary radii associated
  with these solutions are all different.}
\label{fig:lambdaNuR}
\end{figure}

When the parameters are fixed as mentioned, one can look at the
geometric tensors inside the sphere.  First the metric functions
\(\rme^{\lambda}\)
and \(\rme^{\nu}\)
are investigated when the boundary radius in
figure~\ref{fig:lambdaNuR} is changing.

In~\myfigref{fig:lambdaNuR}{fig:lambdaR} the normalization of the
Schwarzschild interior at \(r_{b}=1\)
is clearly seen.  The same is true
in~\myfigref{fig:lambdaNuR}{fig:nuR}, but this particular feature is
harder to see clearly.  All the internal solutions match to the same
exterior Schwarzschild metric by construction, even though they all do
so at different radii.  This is to be expected, with the \(\mu=1\)
case having the maximum radius of
\(r_{b} = \sqrt[3]{5/2} \approx 1.36,\)
since the density shows the greatest transition (from a central value
until it vanishes at the boundary), and to achieve the same mass with
a smoother decrease in density a larger sphere is needed.  The later
case is also the only one where the matching of the metric function
results in the matching of the metric derivatives too: this is seen as
the slope matching in those particular curves, at the point of contact
between interior and exterior.

\begin{figure}[h!]
 \centering
   \begin{subfigure}[b]{0.5\textwidth}
       \includegraphics[width=\textwidth]{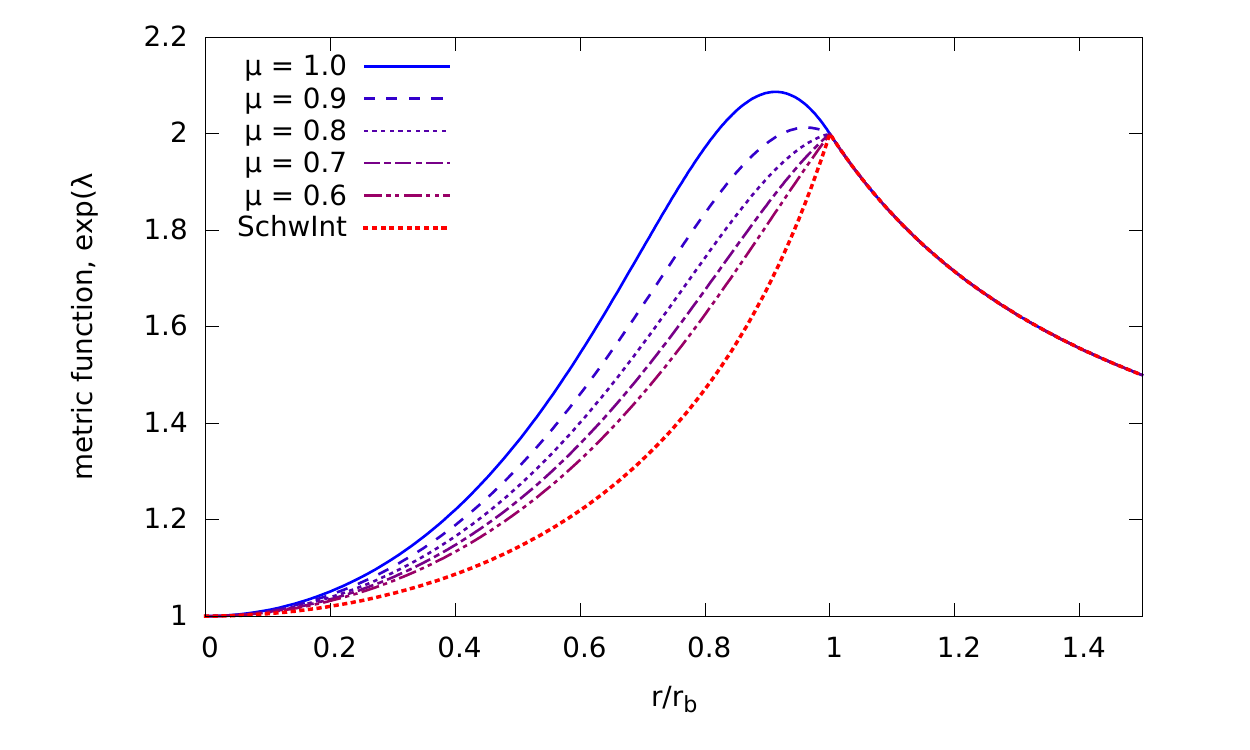}
       \caption{The $\lambda$ metric function}\label{fig:lambdaM}
   \end{subfigure}~
   \begin{subfigure}[b]{0.5\textwidth}
       \includegraphics[width=\textwidth]{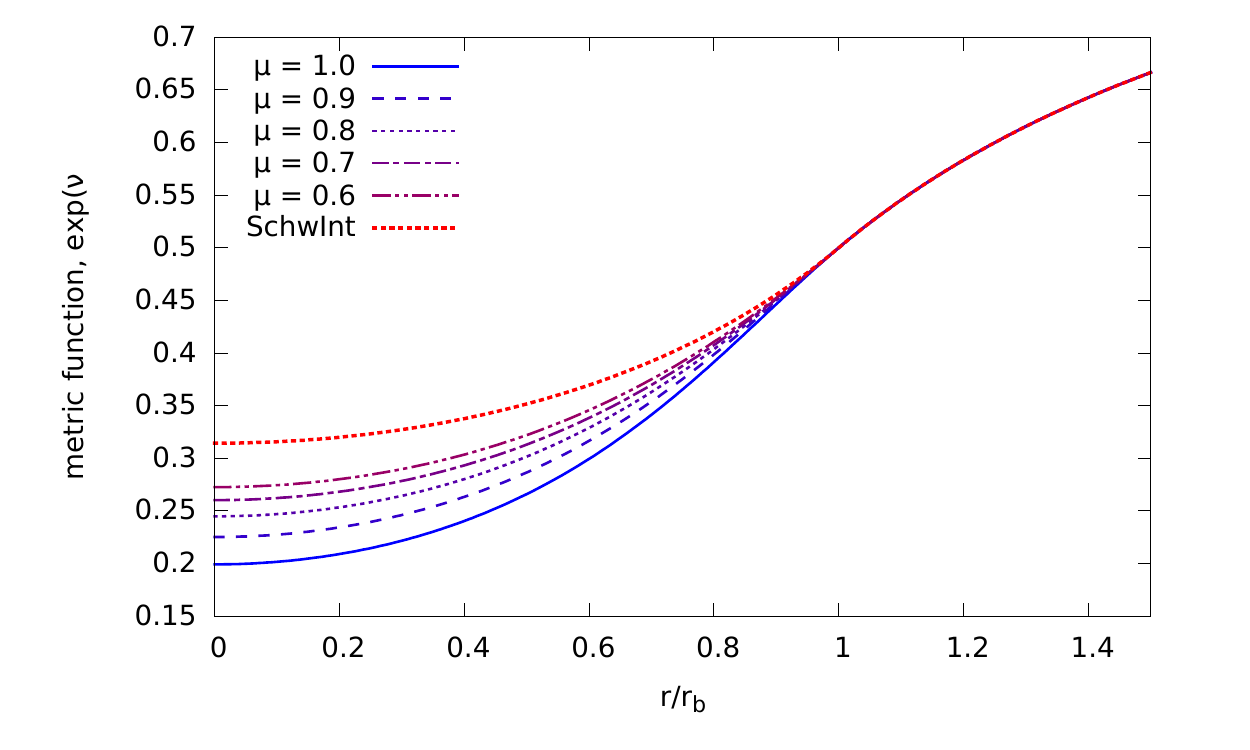}
       \caption{The $\nu$ metric function}\label{fig:nuM}
   \end{subfigure} 
\caption{(Colour online) The metric functions with the radial
  coordinate inside and outside the star for the TVII solution. The
  parameter values are $M = 1/4$ and $\mu$ taking the
  various values shown in the legend.  Also shown for comparison,
  the Schwarzschild interior metric for the same parameter
  values. Again the unmistakable trend of $\lim_{\mu \to 0}$ of TVII
  resulting in Schwarzschild interior is clear. The central density
  $\rho_{c}$ associated with these solutions are all different.}
\label{fig:lambdaNuM}
\end{figure}

If instead one fixes \(r_{b}=1\)
and varies the central density to achieve the same external mass,
Figure~\ref{fig:lambdaNuM} is obtained, and again one sees a similar
pattern as the previous graph sets, with the smooth transition of the
metric again occurring for the \(\mu=1\) curve only.

From these metric functions, all other geometric tensors can be
computed.  However the interpretation of the Riemann tensor's
components directly is difficult since many of the symmetries in this
particular tensor make for multiple components to carrying the same
information.  Instead of analyzing the Riemann tensor \(R_{abcd}\)
components directly, we will use its decomposition into its Weyl
\(C_{abcd},\)
Ricci tensor \(R_{ab} = g^{cd}R_{cadb},\)
and Ricci scalar \(R = g^{ab}R_{ab}\) contributions through
\begin{equation}
  \label{eq:Weyl}
R_{abcd} = C_{abcd}  - g_{a[d}R_{c]b} - g_{b[c}R_{d]a} - \f{1}{3} R g_{a[c}g_{d]b},
\end{equation}
to give a clearer picture of how these encode
gravity together.  It will be evident from these figures that the diagonal
components of the Ricci tensor are non-zero inside the sphere, but
must vanish in the Ricci-flat exterior at the star's boundary.

\subsection{\label{ssec:Ric}Ricci tensor components}
\begin{figure}[h!]
 \centering
   \begin{subfigure}[t]{0.5\linewidth}
       \includegraphics[width=\textwidth]{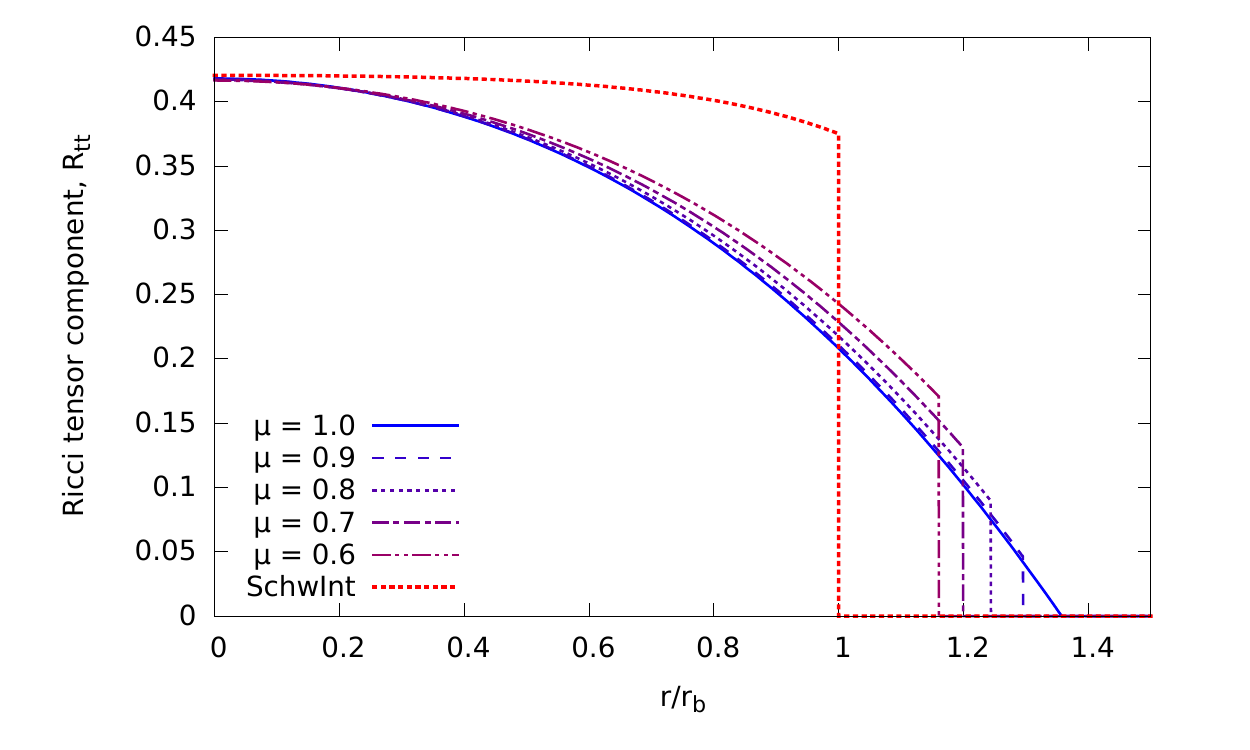}
       \caption{Constant central density}\label{fig:R11R}
   \end{subfigure}~
   \begin{subfigure}[t]{0.5\textwidth}
       \includegraphics[width=\textwidth]{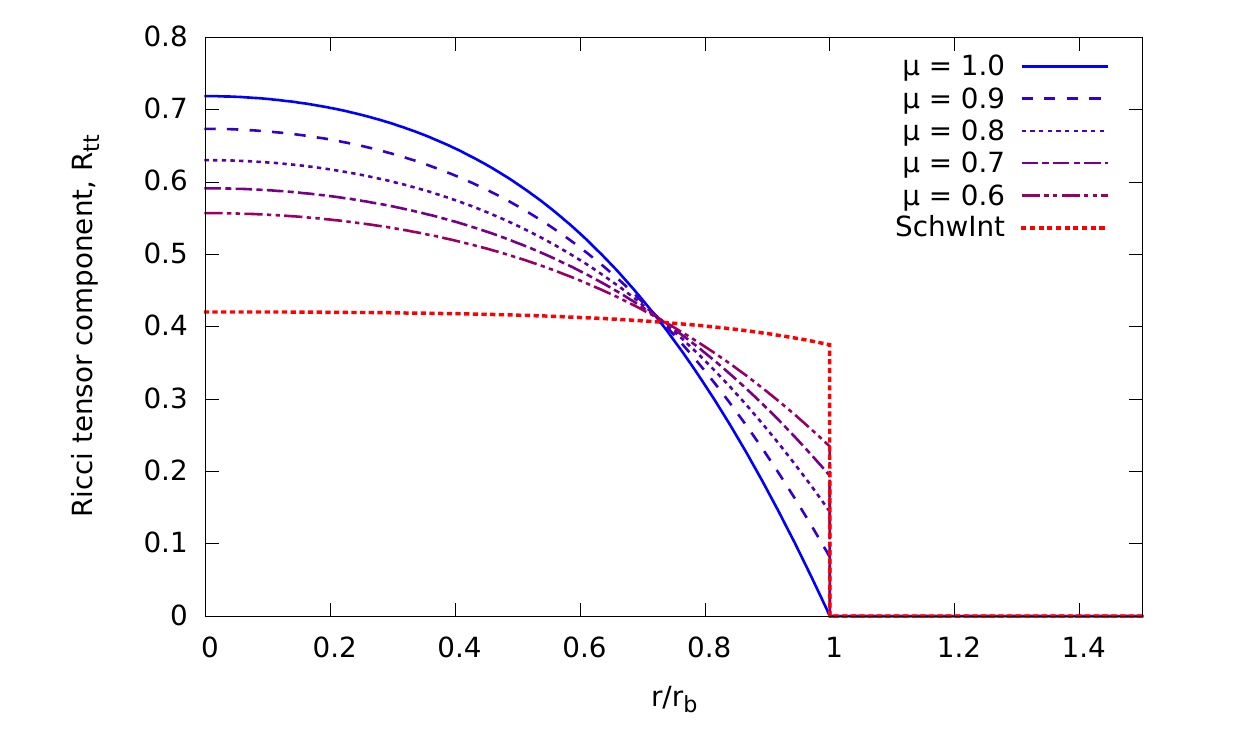}
       \caption{Constant radius}\label{fig:R11M}
   \end{subfigure} 
   \caption{(Colour online) The $R_{tt}$ tensor component with the
     radial coordinate inside and outside the star for the TVII
     solution. The parameter values are
     $M = 1/4, \rho_{c} = 3/(16\pi)$ and $\mu$ taking the various
     values shown in the legend for figure~(\subref{fig:R11R}), and
     $M = 1/4$ for figure~(\subref{fig:R11M}). Also shown
     for comparison, the Schwarzschild interior metric for the same
     parameter values.}
\label{fig:Ric11}
\end{figure}

While the central value of this Ricci component does not change by
much in the first panel of \myfigref{fig:Ric11}{fig:R11R} which plots
the Ricci tensor component \(R_{tt},\)
when \(\mu\)
gets progressively closer to zero, the discontinuity at the boundary
does increase, the continuous curve being the one with \(\mu=1:\)
the ``natural'' case again.  This trend is again seen
in~\myfigref{fig:Ric11}{fig:R11M}, however with the central value
changing too.  The unmistakable trend of $\lim_{\mu \to 0}$ of TVII
resulting in the Schwarzschild interior is clear.

\begin{figure}[h]
   \centering
   \begin{subfigure}[b]{0.5\textwidth}
       \includegraphics[width=\textwidth]{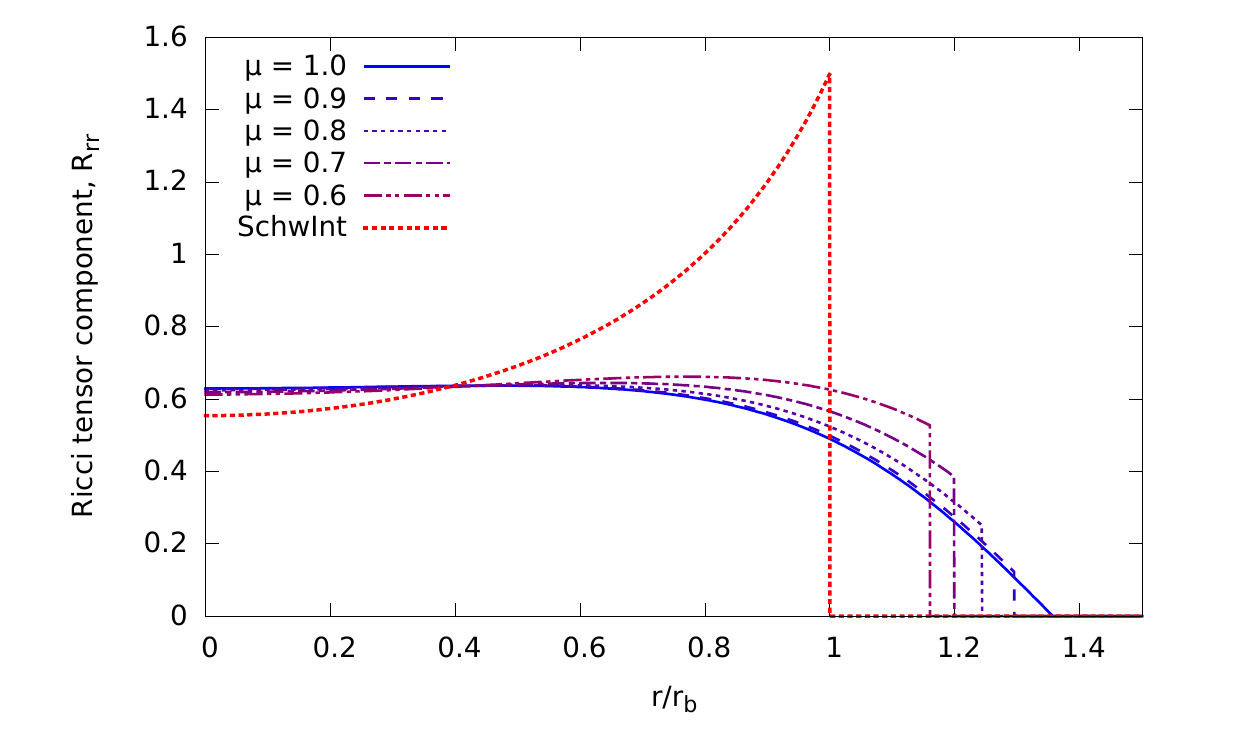}
       \caption{Constant central density}\label{fig:R22R}
   \end{subfigure}~
   \begin{subfigure}[b]{0.5\textwidth}
       \includegraphics[width=\textwidth]{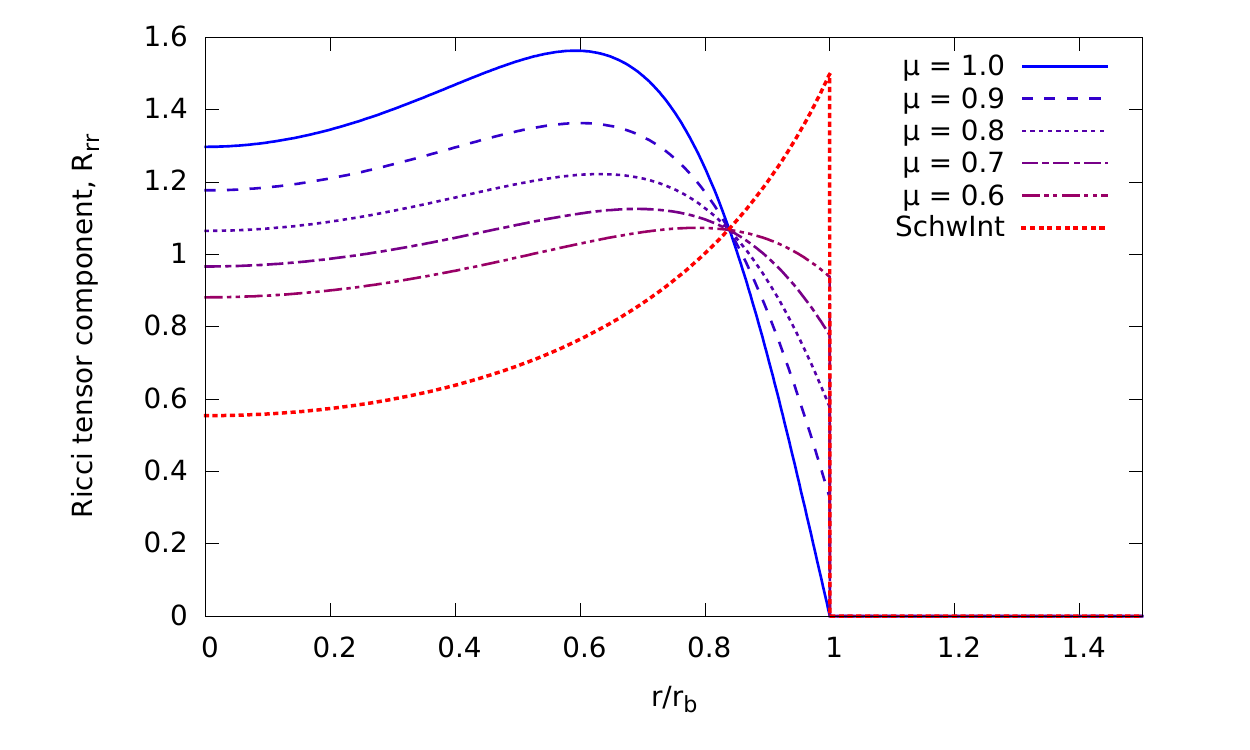}
       \caption{Constant radius}\label{fig:R22M}
   \end{subfigure}
   \caption{(Colour online) The $R_{rr}$ tensor component with the
     radial coordinate inside and outside the star for the TVII
     solution. The parameter values are
     $M = 1/4, \rho_{c} = 3/(16\pi)$ and $\mu$ taking the various
     values shown in the legend for figure(~\subref{fig:R22R}), and
     $M = 1/4$ for figure~(\subref{fig:R22M}).  Also shown
     for comparison is the Schwarzschild interior metric for the same
     parameter values.}
\label{fig:Ric22}
\end{figure}

The second Ricci tensor component: \(R_{rr}\)
is shown next in Figure~\ref{fig:Ric22}, where similar features are
seen.  The largest discontinuity occurs in the Schwarzschild interior
case.  The striking convergence of the curves at around
\(r \approx 0.839\)
in the second panel~\myfigref{fig:Ric22}{fig:R22M} is due to a lack of
higher resolution that would show that the curves indeed cross at
different points.  However the value of \(R_{rr}\) in the exterior
\(r \geq r_{b}\) is identically zero.

\begin{figure}[h]
   \centering
   \begin{subfigure}[b]{0.5\textwidth}
       \includegraphics[width=\textwidth]{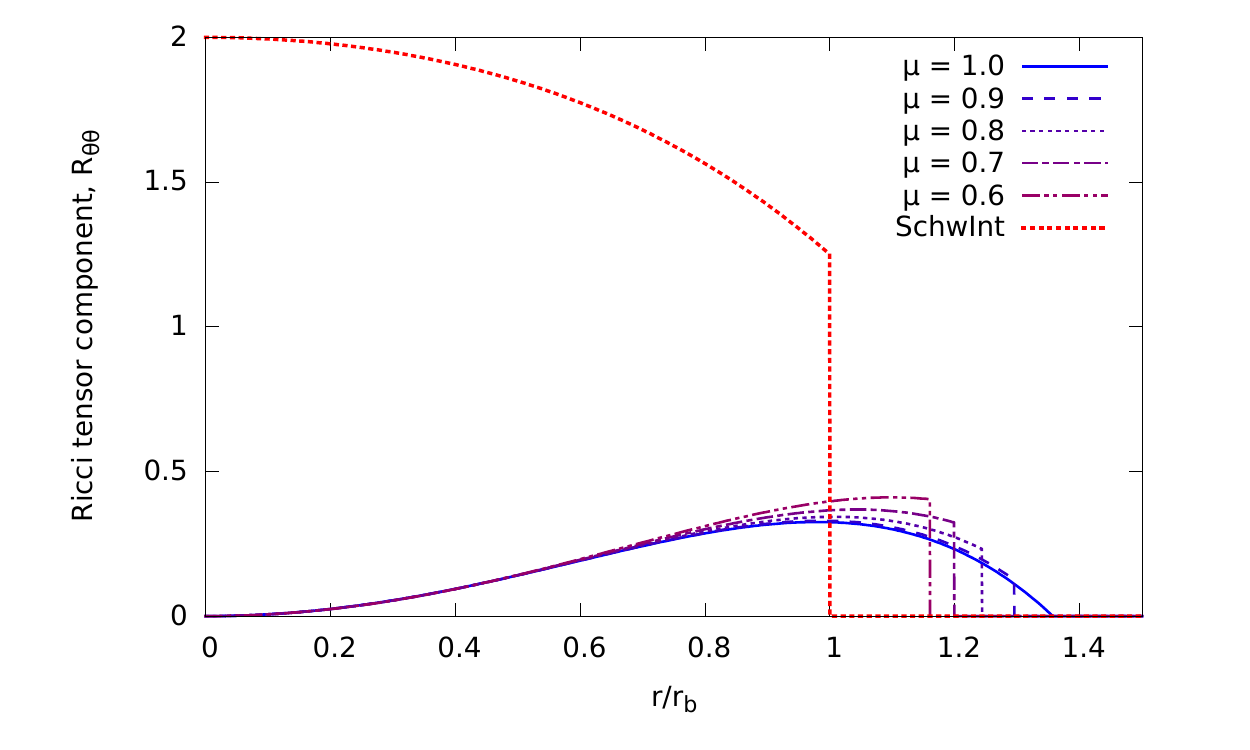}
       \caption{Constant central density}\label{fig:R33R}
   \end{subfigure}~
   \begin{subfigure}[b]{0.5\textwidth}
       \includegraphics[width=\textwidth]{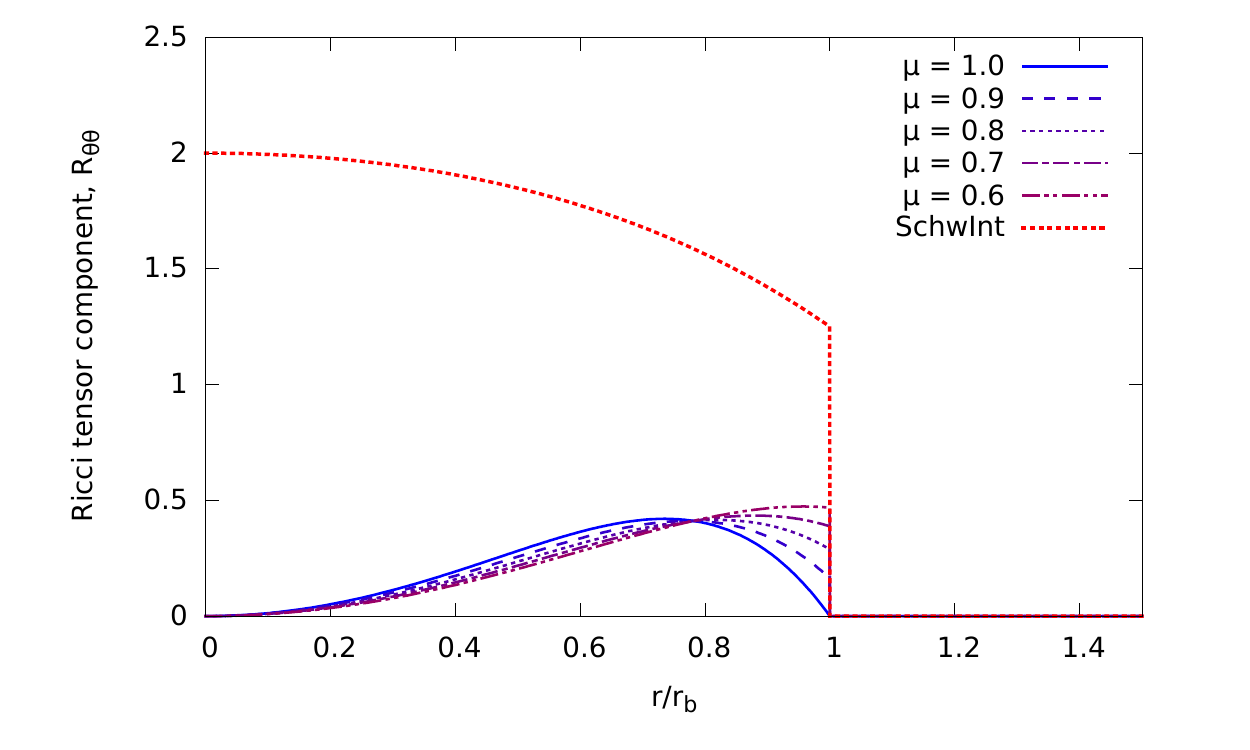}
       \caption{Constant radius}\label{fig:R33M}
   \end{subfigure}
   \caption{(Colour online) The $R_{\theta \theta}$ tensor component
     with the radial coordinate inside and outside the star for the
     TVII solution. The parameter values are
     $M = 1/4, \rho_{c} = 3/(16\pi)$ and $\mu$ taking the various
     values shown in the legend for figure~(\subref{fig:R33R}), and
     $M = 1/4$ for figure~(\subref{fig:R33M}).  Also shown
     for comparison is the Schwarzschild interior metric for the same
     parameter values.}
\label{fig:Ric33}
\end{figure}

Next the \(R_{\theta\theta}\)
component is shown in Figure~\ref{fig:Ric33}.  The unusual feature
here is the completely different shape of the Schwarzschild interior
component of the Ricci tensor.  However an intuitive explanation of
this feature is that since the Weyl tensor components which are shown
next in Figures~\ref{fig:W21}, \ref{fig:W31}, and~\ref{fig:W43} have
to be zero in the Schwarzschild interior solution, the Ricci
components have to compensate by having larger values so that the
Riemann tensor inside is still non-zero.  The convergence of the lines
around \(r \approx 0.8\)
is again an effect of the lower resolution used
in~\myfigref{fig:Ric33}{fig:R33M}

The last diagonal Ricci component \(R_{\varphi \varphi}\)
is the same as \(R_{\theta \theta}\)
modulo a \(\sin(\theta)\)
term, and will not be shown here, instead   
\begin{figure}[h]
   \centering
   \begin{subfigure}[b]{0.5\textwidth}
       \includegraphics[width=\textwidth]{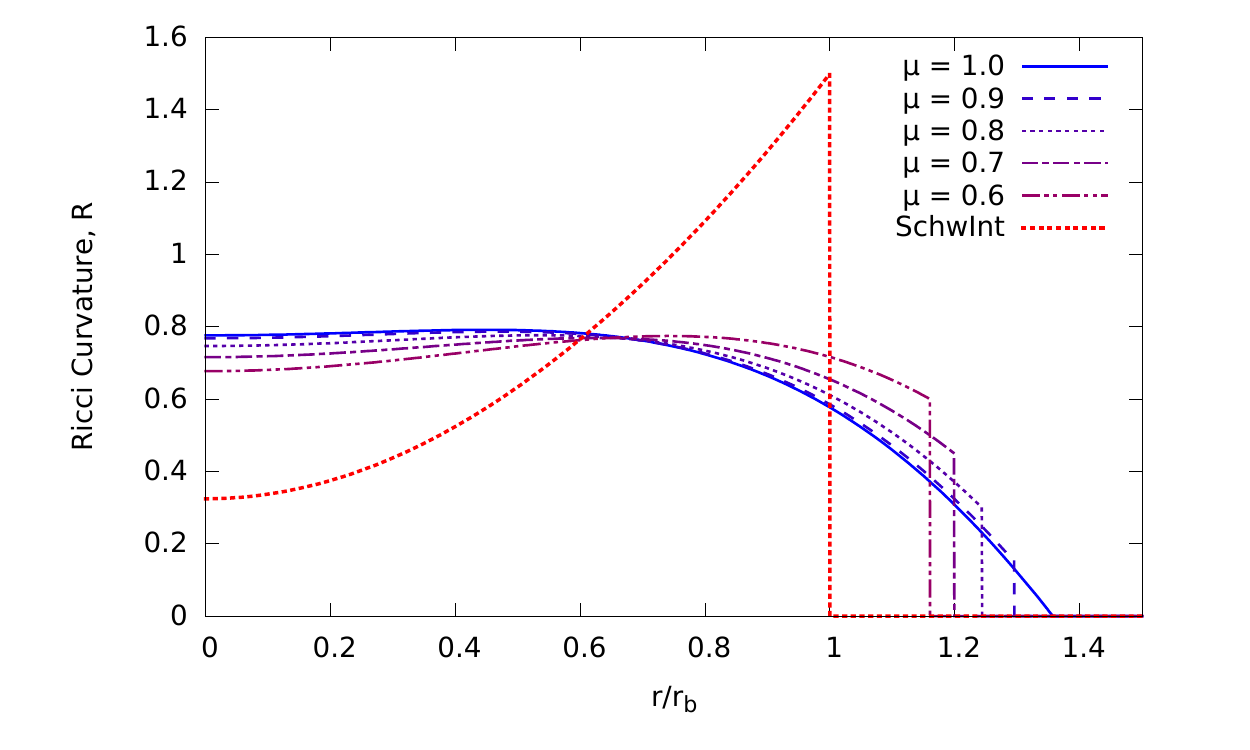}
       \caption{Constant central density}\label{fig:RCR}
   \end{subfigure}~
   \begin{subfigure}[b]{0.5\textwidth}
       \includegraphics[width=\textwidth]{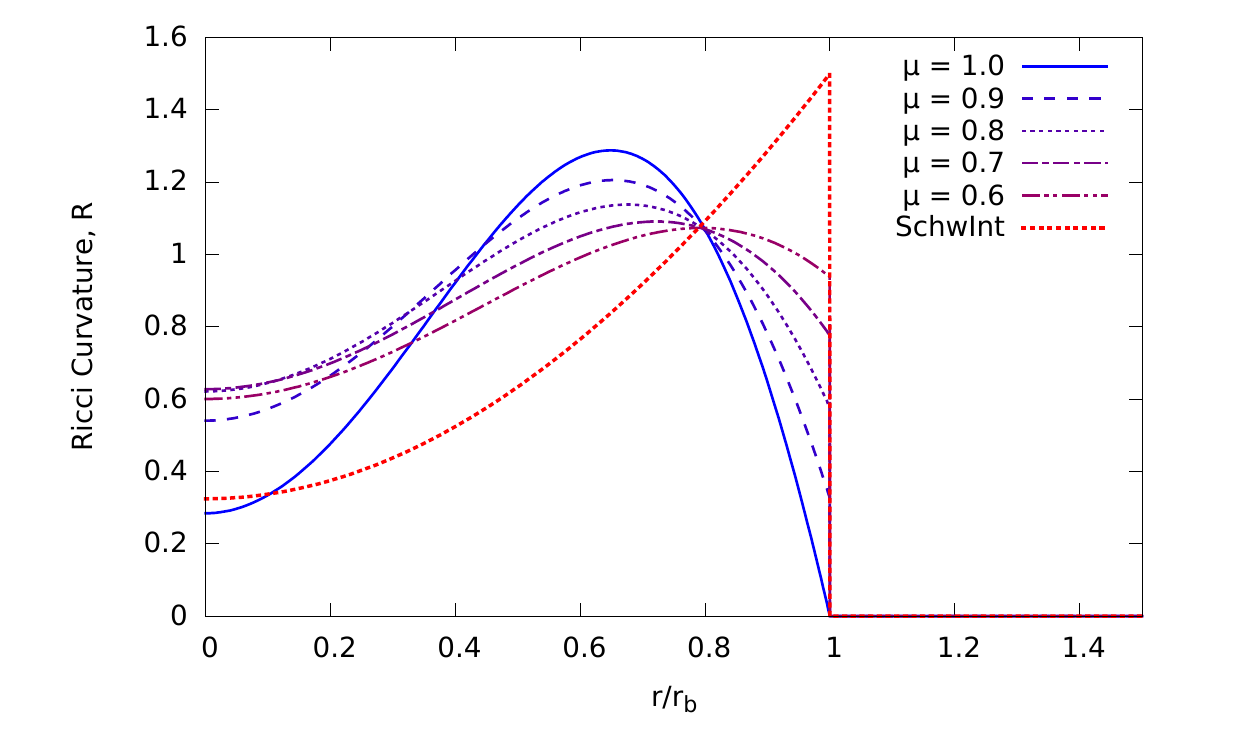}
       \caption{Constant radius}\label{fig:RCM}
   \end{subfigure}
   \caption{(Colour online) The $R$ scalar with the radial coordinate
     inside and outside the star for the TVII solution. The parameter
     values are $M = 1/4, \rho_{c} = 3/(16\pi)$ and $\mu$ taking the
     various values shown in the legend for figure~(\subref{fig:RCR}),
     and $M = 1/4$ for figure~(\subref{fig:RCM}).  Also shown
     for comparison, the Schwarzschild interior metric for the same
     parameter values. The relationship between Tolman~VII and
     Schwarzschild interior is clear in the constant density case, but
     much harder to see in the constant radius case.}
\label{fig:RicCur}
\end{figure}
the Ricci scalar: the
trace of the Ricci tensor, is shown in Figure~\ref{fig:RicCur}.

\subsection{\label{ssec:Wey} Weyl tensor components}
\begin{figure}[h]
   \centering
   \begin{subfigure}[b]{0.5\textwidth}
       \includegraphics[width=\textwidth]{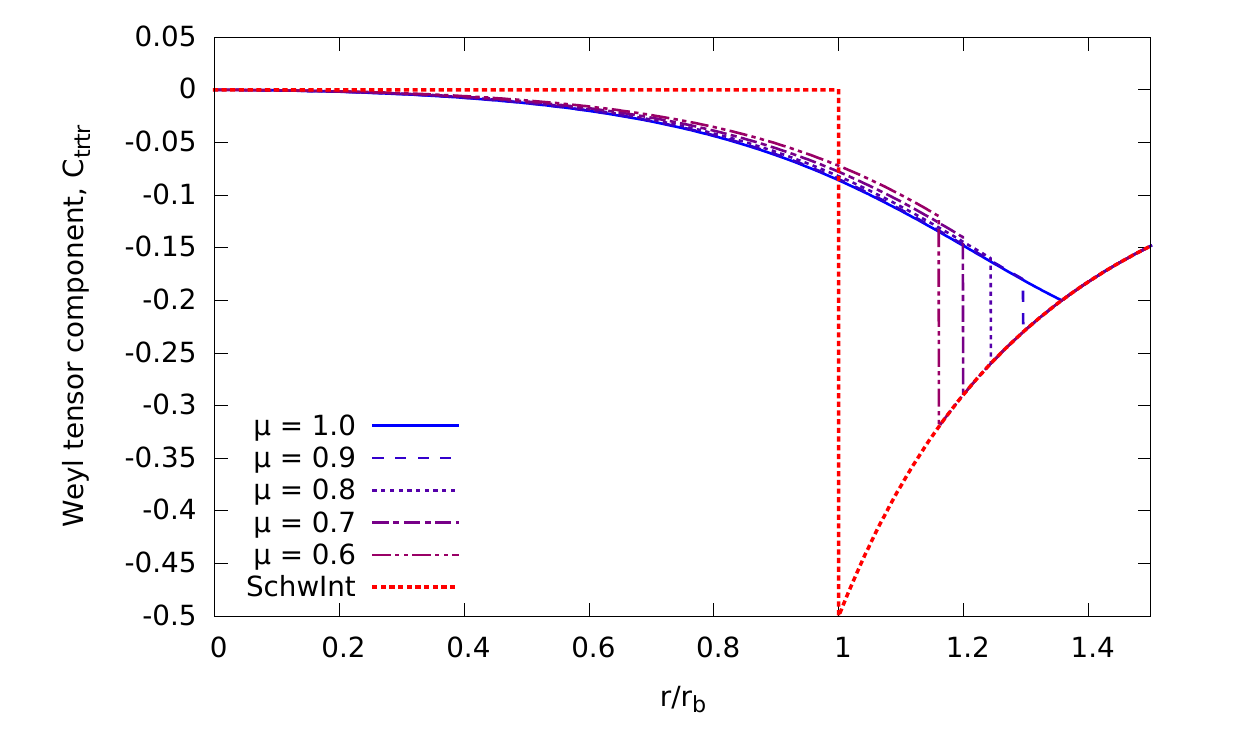}
       \caption{Constant central density}\label{fig:W21R}
   \end{subfigure}~
   \begin{subfigure}[b]{0.5\textwidth}
       \includegraphics[width=\textwidth]{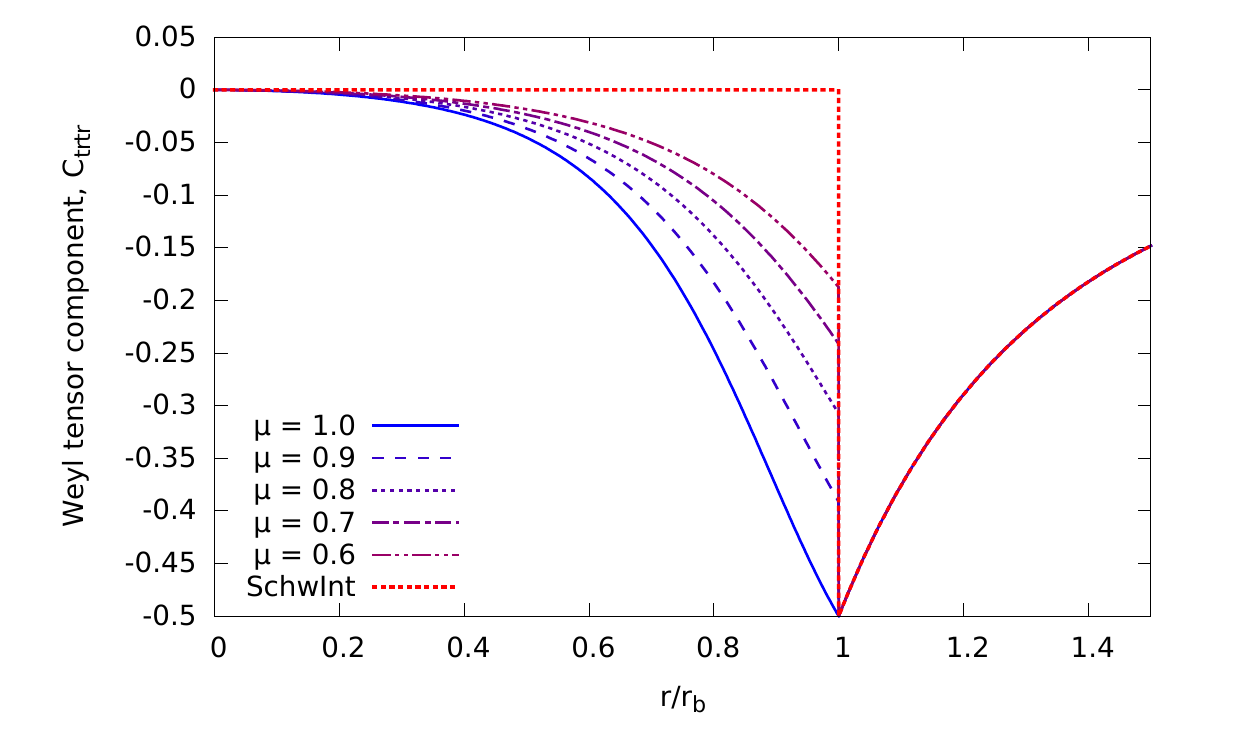}
       \caption{Constant radius}\label{fig:W21M}
   \end{subfigure}
   \caption{(Colour online) The $C_{t r t r}$ tensor component with
     the radial coordinate inside and outside the star for the TVII
     solution. The parameter values are
     $M = 1/4, \rho_{c} = 3/(16\pi)$ and $\mu$ taking the various
     values shown in the legend for figure~(\subref{fig:W21R}), and
     $M = 1/4$ for figure~(\subref{fig:W21M}).  Also shown
     for comparison is the Schwarzschild interior metric for the same
     parameter values.}
\label{fig:W21}
\end{figure}

Next all the non-zero Weyl tensor components are investigated.  While
there are six non-zero components, only three are unique.  The first one
of interest is the \(C_{trtr}\)
component shown in Figure~\ref{fig:W21}.
The \(C_{r \theta r \theta}\)
and the $C_{r \varphi r \varphi}$ tensor components are the same as
\(C_{trtr},\)
and will not be shown separately.  The Weyl tensor components can be
understood in a similar way as the Ricci components.  First one notes
the same trend as noted previously; namely the continuity of the
components in the natural \((\mu=1)\)
case across the fluid boundary.  The discontinuities become more and
more pronounced as \(\mu\)
tends towards zero, with the largest discontinuity occurring in the
Schwarzschild interior case, which is the formal \(\mu \to 0\)
limit.  This trend is also clear in the constant radius case in
~\myfigref{fig:W21}{fig:W21M}.  Contrary to the Ricci case however, since the
Schwarzschild \emph{exterior} solution is a pure Weyl solution, the
tensor components do not vanish in the exterior, and one can see the
\(-2M/r^{3}\)
fall off clearly.  Furthermore the fact the the Schwarzschild interior
is a pure Ricci solution becomes clear: The Weyl tensor components for
this particular solution is identically zero everywhere inside the
star.

\begin{figure}[h]
   \centering
   \begin{subfigure}[b]{0.5\textwidth}
       \includegraphics[width=\textwidth]{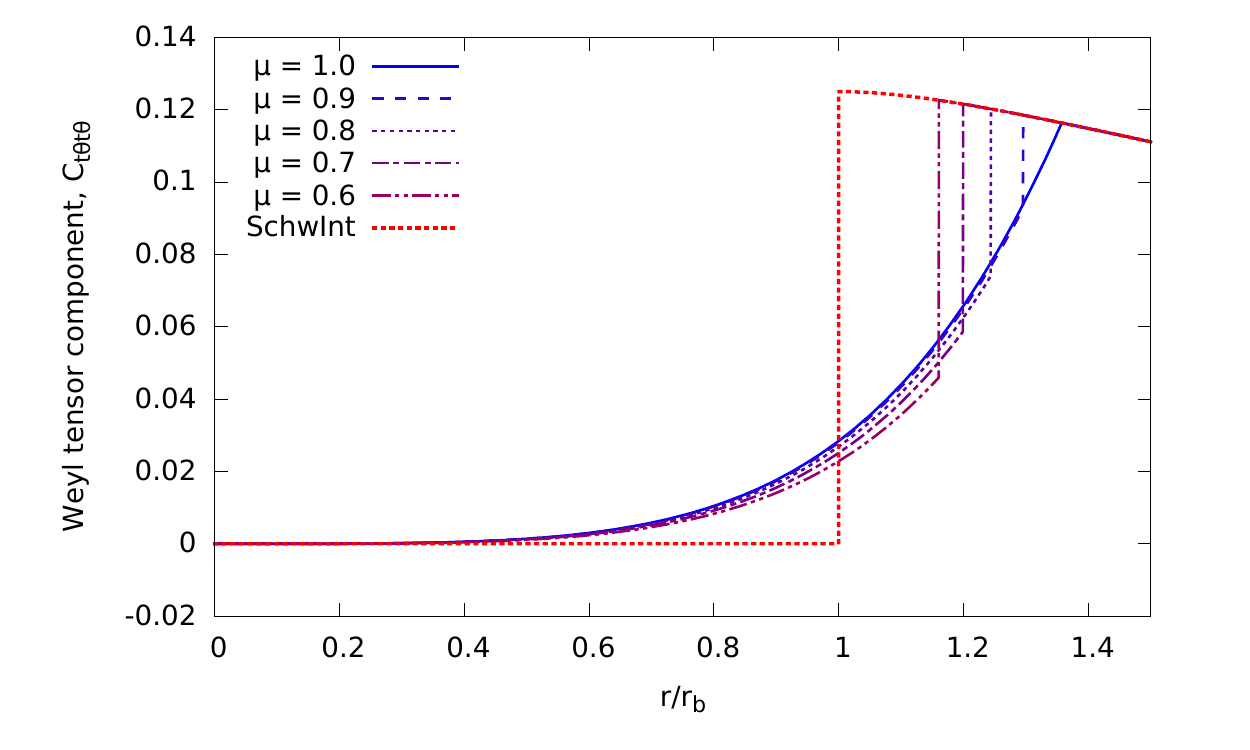}
       \caption{Constant central density}\label{fig:W31R}
   \end{subfigure}~
   \begin{subfigure}[b]{0.5\textwidth}
       \includegraphics[width=\textwidth]{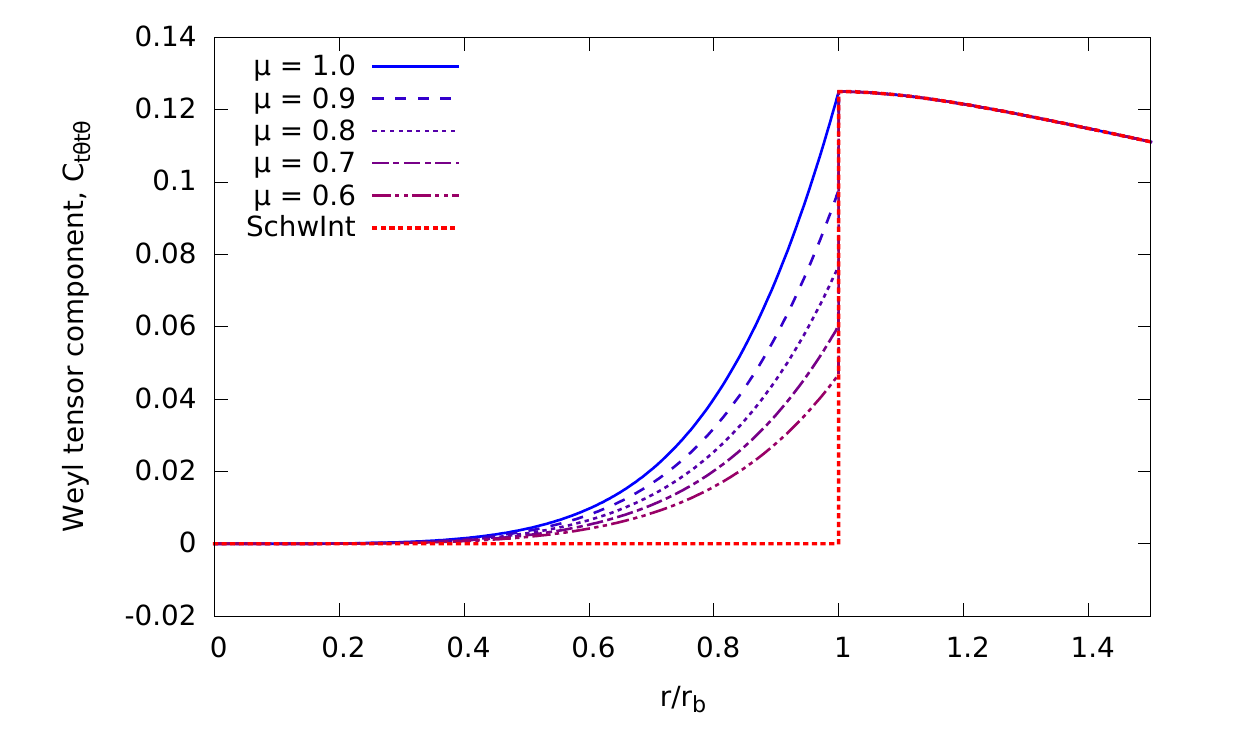}
       \caption{Constant radius}\label{fig:W31M}
   \end{subfigure}
   \caption{(Colour online) The $C_{t \theta t \theta}$ tensor
     component with the radial coordinate inside and outside the star
     for the TVII solution. The parameter values are
     $M = 1/4, \rho_{c} = 3/(16\pi)$ and $\mu$ taking the various
     values shown in the legend for figure~(\subref{fig:W31R}), and
     $M = 1/4$ for figure~(\subref{fig:W31M}).  Also shown
     for comparison is the Schwarzschild interior metric for the same
     parameter values.}
\label{fig:W31}
\end{figure}

Next the \(C_{t \theta t \theta}\)
component is shown in Figure~\ref{fig:W31}.  The main trends seen previously
are still present in these, and the only drastically different
behaviour is the slower fall off of the Weyl tensor component as
compared to the previous one in the exterior.  This follows from the
more complicated expression of that components which falls off as
\(1/r\)
for large \(r\)
in this case.  The \(C_{t \varphi t \varphi}\)
behaviour is the same as that for \(C_{t \theta t \theta},\) and is not shown separately.

\begin{figure}[h]
   \centering
   \begin{subfigure}[b]{0.5\textwidth}
       \includegraphics[width=\textwidth]{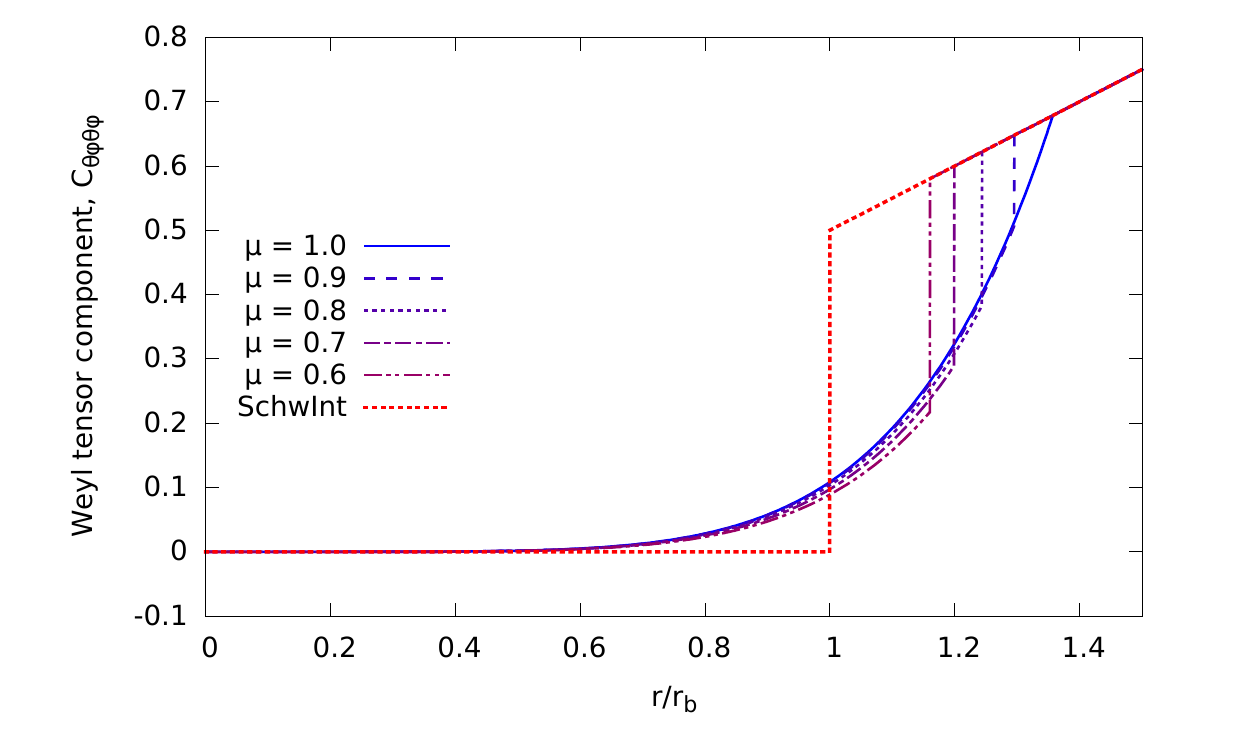}
       \caption{Constant central density}\label{fig:W43R}
   \end{subfigure}~
   \begin{subfigure}[b]{0.5\textwidth}
       \includegraphics[width=\textwidth]{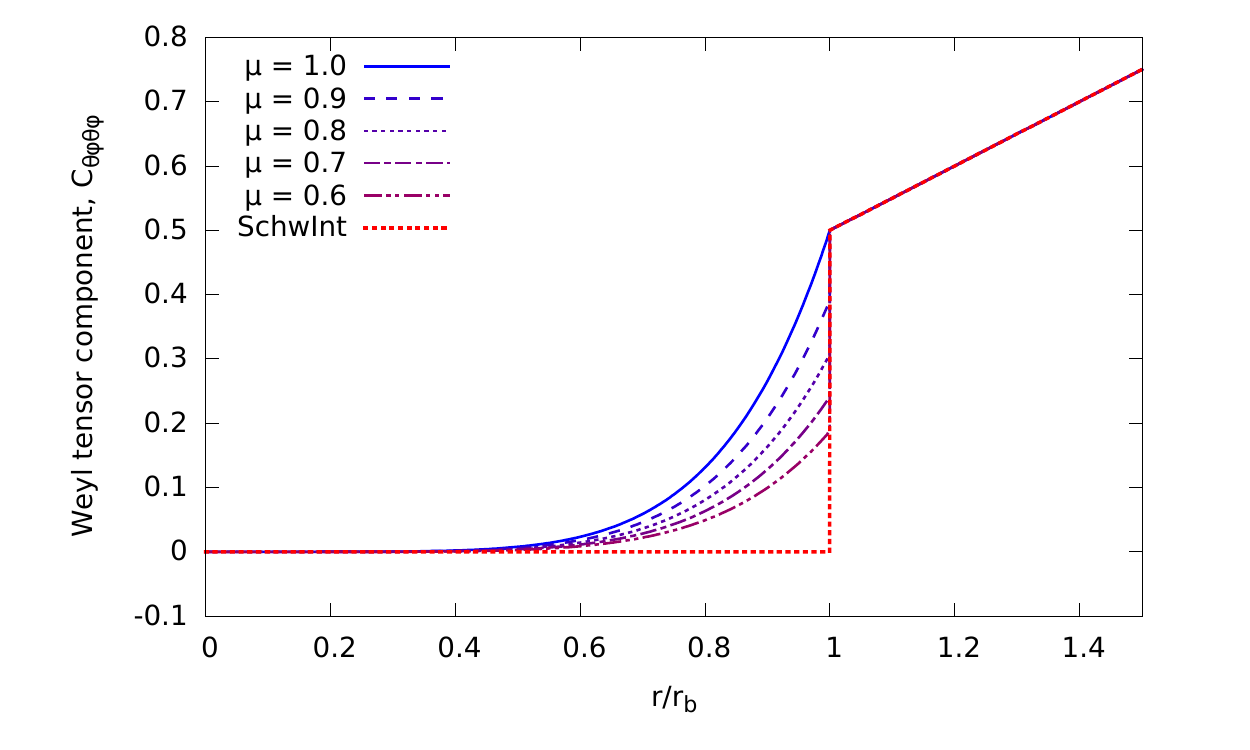}
       \caption{Constant radius}\label{fig:W43M}
   \end{subfigure}
   \caption{(Colour online) The $C_{\theta \varphi \theta \varphi}$
     tensor component with the radial coordinate inside and outside
     the star for the TVII solution. The parameter values are
     $M = 1/4, \rho_{c} = 3/(16\pi)$ and $\mu$ taking the various
     values shown in the legend for figure~(\subref{fig:W43R}), and
     $M = 1/4$ for figure~(\subref{fig:W43M}).  Also shown
     for comparison is the Schwarzschild interior metric for the same
     parameter values.}
\label{fig:W43}
\end{figure}

The \(C_{\theta \varphi \theta \varphi}\)
component however is different and is shown in Figure~\ref{fig:W43}.
The increasing Weyl tensor component in the exterior is surprising, as
it would be expected that the Weyl tensor should vanish at infinity.
However this is only an artifact of using the fully covariant tensor
\emph{components} instead of \(C^{a}{}_{bcd},\)
which is more readily interpreted in a physical context: indeed, the
latter form of the tensor components show no such increase to infinity
for \(r \to \infty.\)
The fully covariant components were used in this article because they
made subsequent calculations simpler.

This ends the presentation of the Weyl tensor components.  All of them
show the same general trends relating the \(\mu\)
values from the Tolman~VII solution to the Schwarzschild solution.
The continuity of the \(\mu=1,\)
``natural'' case is also readily apparent, and the larger
discontinuities in the tensor components with decreasing \(\mu\)
up to the limiting Schwarzschild interior solution which needs zero
interior Weyl component, but non-zero exterior values is seen in all
the non identically zero components.

\section{\label{sec:tetrads}Tetrad formalisms}
Another equivalent way of looking at the tensor components, but in a
coordinate invariant way is through the use of the tetrad formalisms
such as the Newman-Penrose (NP) formalism which uses a set of four
null vectors to form a frame (vierbein), or the any other non-unique
orthonormal set of four vectors to form the frame.  Both methods will
be investigated next, while keeping in mind that these are but
different ways of looking at the same solution.  The advantage of
a tetrad formulation is that it provides definitions of a number of scalars.  The scalars encode the same information as the tensor components investigated in
the previous sections, without using the Einstein summation
convention, while at the same time allowing the study of asymptotic
behaviour of the curvature quantities~\cite{NewTod80}.

\subsection{\label{ssec:tetradOrtho}An orthogonal tetrad}
One proceeds by picking a frame through the definition of four
orthonormal basis vectors
\begin{subequations}
  \label{eq:NTetrad}
  \begin{align}
    \vec{e}_{(0)}{}^{a} &= \vec{v}^{a} = \left( \f{1}{c_{1}\cos{(\phi \xi)} + c_{2} \sin{(\phi \xi)}}, 0 ,0,0 \right) \\
    \vec{e}_{(1)}{}^{a} &= \vec{i}^{a} = \left( 0, -\sqrt{ar^{4} - br^{2} + 1},0,0 \right) \\
    \vec{e}_{(2)}{}^{a} &= \vec{j}^{a} =  \left( 0,0,-\f{1}{r},0 \right) \\
    \vec{e}_{(3)}{}^{a} &= \vec{k}^{a} =  \left( 0,0,0,-\f{1}{r \sin{\theta}} \right)
  \end{align}
\end{subequations}
to define the orthonormal tetrad which will later be used to define a
null tetrad for the NP-formalism.  It should be mentioned that the
notation for the different vectors and scalars calculated from them
differ in most books, and here the notation used in the differential
geometry package of Maple\texttrademark{} is what is being followed.
The notation used in~\cite{NewTod80} and in~\cite{Cha98} write
\(R_{ab}\)
for \(R_{(a)(b)}\)
so that \(R_{(1)(1)}\)
becomes \(R_{11}\) for example, and this can lead to confusion.
\subsection{\label{ssec:tetradNull}The NP tetrad}
In the NP formalism, The
null tetrad used in the usual notation are
\begin{subequations}
  \label{eq:NPTetrad}
  \begin{align}
    \vec{e}_{(0)}{}^{a} &= \vec{l}^{a} = \f{1}{\sqrt{2}} \left( \f{1}{c_{1}\cos{(\phi \xi)} + c_{2} \sin{(\phi \xi)}}, -\sqrt{ar^{4} - br^{2} + 1},0,0 \right) \\
    \vec{e}_{(1)}{}^{a} &= \vec{n}^{a} = \f{1}{\sqrt{2}} \left( \f{1}{c_{1}\cos{(\phi \xi)} + c_{2} \sin{(\phi \xi)}}, \sqrt{ar^{4} - br^{2} + 1},0,0 \right) \\
    \vec{e}_{(2)}{}^{a} &= \vec{m}^{a} = \f{1}{\sqrt{2}} \left( 0,0,-\f{1}{r},-\f{i}{r \sin{\theta}} \right) \\
    \vec{e}_{(3)}{}^{a} &= \vec{\bar{m}}^{a} = \f{1}{\sqrt{2}} \left( 0,0,-\f{1}{r},\f{i}{r \sin{\theta}} \right)
  \end{align}
\end{subequations}
Where \(c_{1}, c_{2}, \phi, \xi, a,\)
and \(b\)
have been given previously in equations~\eqref{eq:c1,c2},
\eqref{eq:Z}, \eqref{eq:Y}, and~\eqref{eq:xiCoth}.  This null tetrad
can easily be obtained from the orthonormal basis
above~\eqref{eq:NTetrad} through a well defined algorithm.  From this
null tetrad the spin coefficients and scalars can be calculated and
plotted for this particular solution.  Being in the static and
spherically symmetric regime, the only Weyl scalar that does not
vanish in this formalism will be \(\Psi_{2},\)
and the some of the Ricci scalars and rotation coefficients will also
vanish because of the symmetry.  Expressions and graphs of the
non-vanishing scalars will be given in the appendix, and here
only a few attributes of these scalars will be discussed.

The Ricci Scalars \(\Phi_{00}\)
and \(\Phi_{11},\)
are related very simply through \(\Phi_{00} = 2\Phi_{11},\)
because of the isotropic nature of the energy-momentum tensor of this
solution.  Similarly the static and spherically symmetric nature of
the solution manifests itself in the vanishing of all the Weyl scalars
except for \(\Psi_{2}.\)
It should therefore come to no surprise that \(\Psi_{2},\)
as the only Weyl scalar component, is related to Ponce~de~Leon's Weyl
function \(W,\)
introduced later in equation~\eqref{eq:W}.  Most of the spin
coefficients vanish, and according to~\cite{NewTod80}, the vanishing
of \(\kappa^{\np},\)
a condition present in this solution, implies that the integral curve
(congruence) of \(\vec{l}_{a} = g_{ab} \vec{l}^{a},\)
is a geodesic.  From the expression of \(\vec{l}^{a}\)
in~\eqref{eq:NPTetrad}, and the form of the metric, this can be easily
ascertained.  Furthermore \(\sigma^{\np} = 0,\)
means that the congruence does not undergo any shear with increasing
\(r\),
and \(\rho^{\np}\)
being non-zero suggests that the congruence expands with increasing
\(r.\) \cite{NewPen62}

\section{\label{sec:discussion}Discussion}
The effects of having a non-zero density at the boundary(which is an
equivalent way of saying that \(\mu \neq 1,\))
and the relation of this effect on the Weyl tensor components had
previously been discussed in~\cite{Pon88} and~\cite{Rag09}.  Of particular
note is the fact that all the Weyl tensor components are related to a
scalar function called \(W\)
in~\cite{Pon88}, and which is defined through
\begin{equation}
  \label{eq:W}
W(r) = \f{r^{2}Z'(Y -rY') + 2rZ ( rY'- Y -r^{2}Y'') + 2rY }{12Y} \xrightarrow{\text{Tolman VII}} 
\f{\kappa \mu \rho_{c}}{15r_{b}^{2}} r^{5},
\end{equation}
in the notation of this article.  The last expression makes it clear
that the value of \(W(r_{b})\)
depends crucially on the value of \(\mu,\)
and indeed on the other parameters too.  As a result the dependence on
\(\mu\)
is becomes clear that in the Schwarzschild interior solution, where
\(\mu\) is zero, \(W\) is zero too.

\begin{figure}[h]
   \centering
   \begin{subfigure}[b]{0.5\textwidth}
       \includegraphics[width=\textwidth]{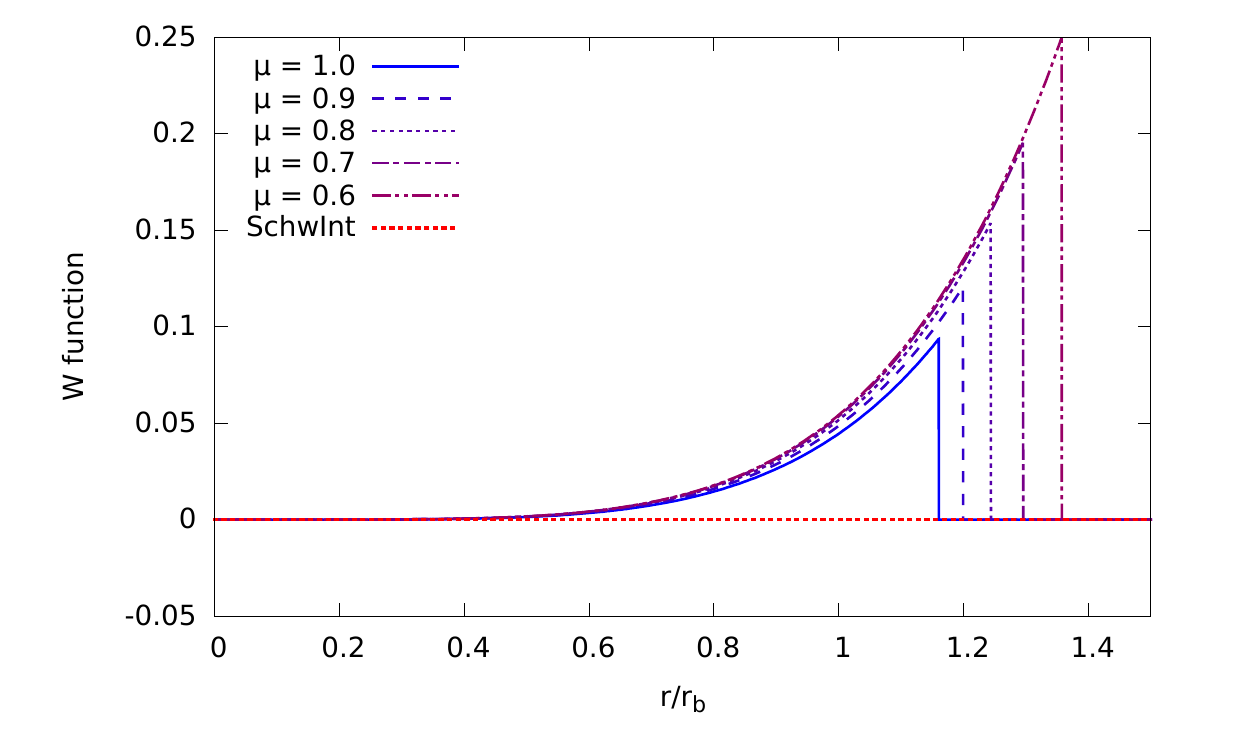}
       \caption{Constant central density}\label{fig:PdLWR}
   \end{subfigure}~
   \begin{subfigure}[b]{0.5\textwidth}
       \includegraphics[width=\textwidth]{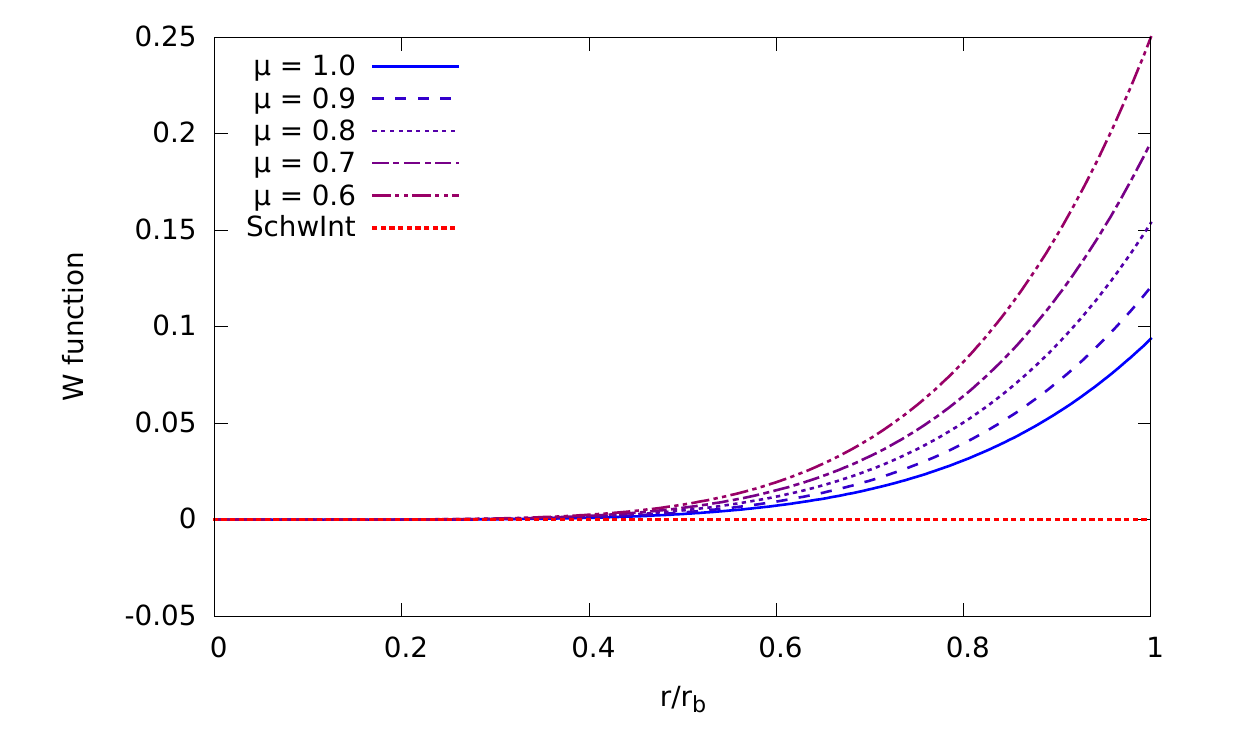}
       \caption{Constant radius}\label{fig:PdLWM}
   \end{subfigure}
   \caption{(Colour online) The $W$ scalar function with the radial
     coordinate inside and outside the star for the TVII solution. The
     parameter values are $M = 1/4, \rho_{c} = 3/(16\pi)$ and $\mu$
     taking the various values shown in the legend for
     figure~(\subref{fig:PdLWR}), and $M = 1/4, r_{b} = 1$ for
     figure~(\subref{fig:PdLWM}). Also shown for comparison is the
     Schwarzschild interior metric for the same parameter values.}
\label{fig:PdLW}
\end{figure}

Ponce~de~Leon~\cite{Pon88} also expressed the
total mass inside the star as a function of \(W\) above as
\begin{equation}
  \label{eq:massW}
m(r)  =\f{4\pi\rho_{c}r^{3}}{3} + W(r).
\end{equation}
This equation can be interpreted in two ways:
\begin{enumerate*}
\item with the Tolman~VII solution seen as a generalization of the
  Schwarzschild interior solution, this relation suggests that the
  external perceived Schwarzschild mass is the exact same as the mass
  due to a sphere of \emph{constant} density, where the central
  density of the Tolman~VII solution determines the Schwarzschild
  interior density.  This contribution corresponds to the first right
  hand term in the mass. Additionally a correction factor that on
  increasing the value of \(\mu\)
  adds more and more mass to the sphere by reducing the boundary
  density discontinuity, achieving zero boundary density at the
  maximum value of \(W,\) is present.
\item However one could also interpret this relation as a definition
  of \(W,\)
  and seen in this light \(W\)
  is measuring the contribution to the ``free'' gravitational field,
  or the free gravitational energy to the externally observed mass
  \(M = m(r_{b}).\)
  This interpretation is further motivated by the
  realization~\cite{Pon88} that
  \(W(r) = rC^{\phi}{}_{\theta \phi \theta},\)
  exactly the same form of equation as
  \(m(r) = rR^{\phi}{}_{\theta \phi \theta},\)
  furthering the idea that some gravitational aspects of the field
  that is similar to mass is being encapsulated by \(W.\)
\end{enumerate*}
The usefulness of \(W,\)
being clear, plots of this function in the constant central densities,
and constant boundary radii cases are given in Figure~\ref{fig:PdLW}.

It is immediately noticed that this function is always positive, and
again according to~\cite{Pon88}, this is to be expected since \(W\)
is only negative in cases where anisotropic pressures exist: an issue
that will be pursued in a later article.  Furthermore, the Buchdahl
limit of \(M/r_{b} = 4/9,\)
cannot be exceeded with the Tolman~VII solution for any parameter
values, as exceeding this limit can only occur if \(W<0.\)
This concludes a geometrical overview of the Tolman~VII solution.

\appendix
\label{A:a}
\section{NP-formalism scalars for the Tolman~VII solution}
All the NP spin coefficients are listed next. The standard notation
according to~\cite{NewPen62, LanLif80} is used throughout, but since
some of the same symbols have already been used in the analysis of the
Tolman~VII solution, the NP coefficients are superscribed to differentiate
them from the constants already used.  First to the coefficients
that vanish:
\begin{equation}
\lambda^{\np}  = \tau^{\np} = \kappa^{\np} = \sigma^{\np} = \pi^{\np} = \nu^{\np} = 0.
\end{equation}
The remaining non-vanishing ones which do not depend on either \(Y\)
or \(Z\) are,
\begin{equation}
\beta^{\np} = -\alpha^{\np} = -\f{\cot{\theta}}{2\sqrt{2} r},  
\end{equation}
as expected from the spherical symmetry of this solution.  These are
the usual spherical coordinates spin coefficient, and they will not be
analyzed any further.

The other spin coefficients which are in terms of the \(Z\)
metric coefficient only are,
\begin{equation}
\label{eq:NPmu}
\mu^{\np} = \rho^{\np} = \f{1}{r} \sqrt{\f{\kappa\mu\rho_{c} r^{4} - 5\kappa\rho_{c}r_{b}^{2} r^{2} + 15 r_{b}^{2}}{30r^{2}_{b}}},  
\end{equation}
and they are plotted next for clarity and completeness in
Figure~\ref{fig:NPmu,rho}.  Since there is an asymptote at \(r=0,\)
as is evident in the expression~\eqref{eq:NPmu},
the plots in Figure~\ref{fig:NPmu,rho} show a ``corrected'' spin
coefficient given by \(r \times \mu^{\np},\)
so that the behaviour for small \(r\)
can also be shown in the same figure.  For \(\mu = 1, \mu^{\np}\)
is continuous as is its radial derivative at \(r=r_{b}.\)
\begin{figure}[h]
   \centering
   \begin{subfigure}[b]{0.5\textwidth}
       \includegraphics[width=\textwidth]{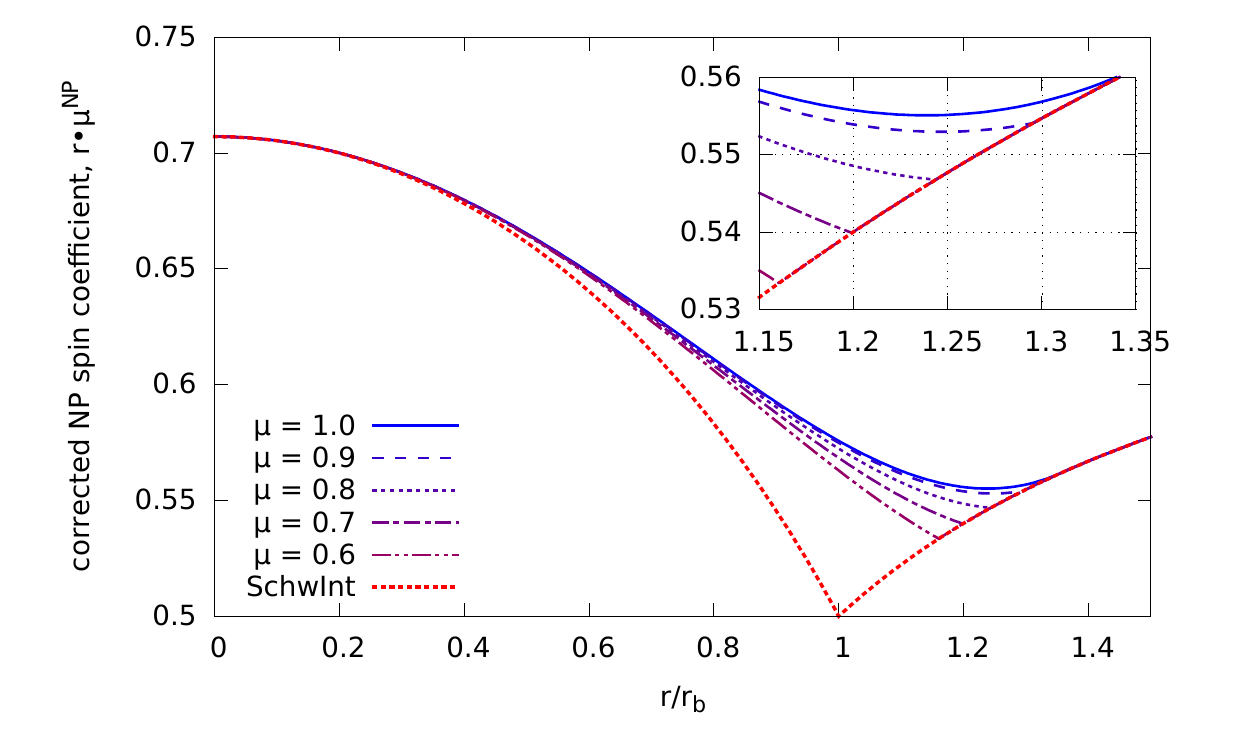}
       \caption{Constant central density}\label{fig:NPmuR}
   \end{subfigure}~
   \begin{subfigure}[b]{0.5\textwidth}
       \includegraphics[width=\textwidth]{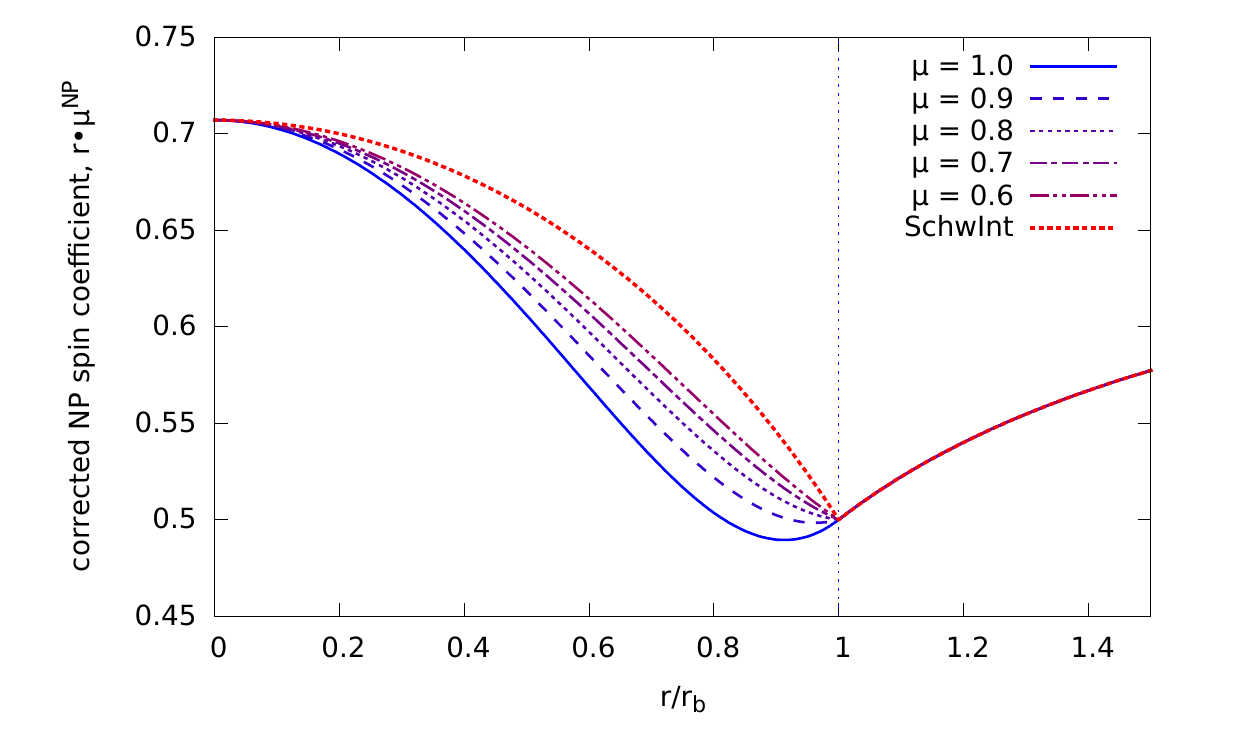}
       \caption{Constant radius}\label{fig:NPmuM}
   \end{subfigure}
   \caption{(Colour online) The ``corrected'' $\mu^{\np}$ spin
     coefficient with the radial coordinate. The parameter values are
     $M = 1/4, \rho_{c} = 3/(16\pi)$ and $\mu$ taking the various
     values shown in the legend for figure~(\subref{fig:PdLWR}), and
     $M = 1/4$ for figure~(\subref{fig:PdLWM}). Also shown for
     comparison is the Schwarzschild interior metric for the same
     parameter values.}
\label{fig:NPmu,rho}
\end{figure}
Finally the remaining spin coefficients which consist of functions of both \(Y\) and \(Z\) are
\begin{equation}
\gamma^{\np} = \epsilon^{\np} = -\f{1}{2} \sqrt{\f{\kappa\mu\rho_{c} r^{4} - 5\kappa\rho_{c}r_{b}^{2} r^{2} + 15 r_{b}^{2}}{30r^{2}_{b}}} \left\{\deriv{}{r} \log{\left[c_{1} \cos (\phi \xi) + c_{2} \sin (\phi \xi)\right]} \right\},  
\end{equation}
where all the constants used in the above have been specified before
in terms of the three \(\mu, \rho_{c}\)
and \(r_{b},\)
and \(\xi\)
has been used as an abbreviated form for \(\xi(r)\)
given in equation~\eqref{eq:xiCoth}.  The above
\(\gamma^{\np}\) is shown in Figure~\ref{fig:NPgamma,epsilon}.
\begin{figure}[h]
   \centering
   \begin{subfigure}[b]{0.5\textwidth}
       \includegraphics[width=\textwidth]{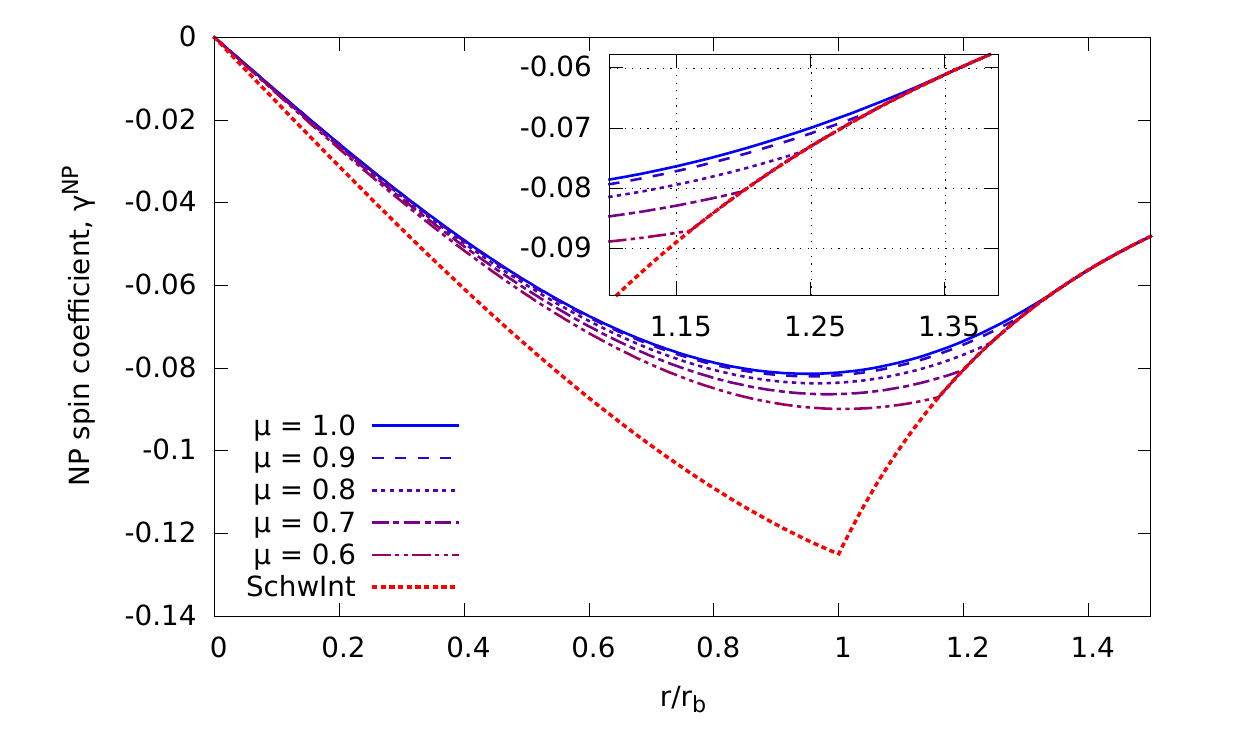}
       \caption{Constant central density}\label{fig:NPgammaR}
   \end{subfigure}~
   \begin{subfigure}[b]{0.5\textwidth}
       \includegraphics[width=\textwidth]{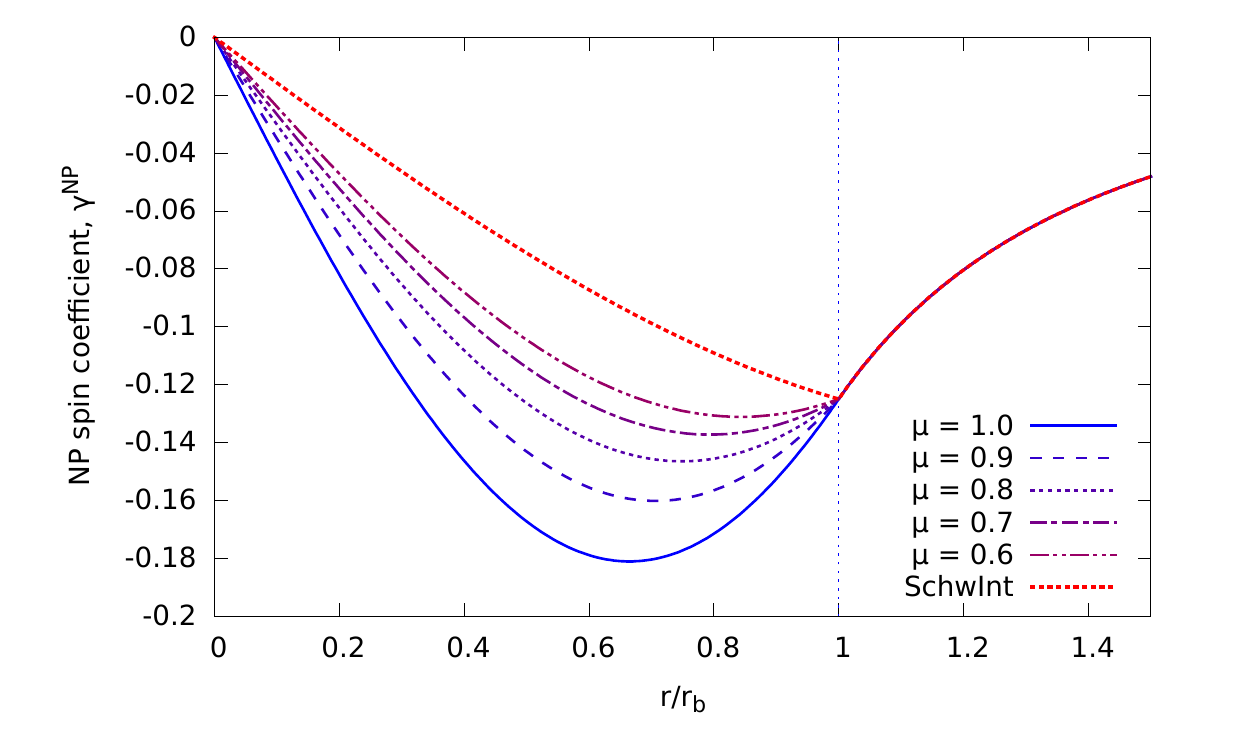}
       \caption{Constant radius}\label{fig:NPgammaM}
   \end{subfigure}
   \caption{(Colour online) The $\gamma^{\np}$ spin coefficient with the
     radial coordinate. The parameter values are
     $M = 1/4, \rho_{c} = 3/(16\pi)$ and $\mu$ taking the various
     values shown in the legend for figure(~\subref{fig:PdLWR}), and
     $M = 1/4$ for figure~(\subref{fig:PdLWM}).  Also shown
     for comparison is the Schwarzschild interior metric for the same
     parameter values.}
\label{fig:NPgamma,epsilon}
\end{figure}

Most of the the Weyl scalars vanish because of the spherically symmetric and
static nature of the solution so that \begin{equation} \Psi_{0}^{\np}
  = \Psi_{1}^{\np}= \Psi_{3}^{\np} = \Psi_{4}^{\np}= 0,
\label{eq:psi013}
\end{equation}
except for \(\Psi_{2}^{\np},\)
which simplifies considerably after a lot of algebra to the expression
\begin{equation}
  \label{eq:psi2}
\Psi_{2}^{\np} = - \f{\kappa \mu \rho_{c}}{15 r^{2}_{b}} r^{2}.
\end{equation}
This is very reminiscent of the function \(W\)
which also simplified to the expression~\eqref{eq:W}.  One can
conclude that indeed Ponce~de~Leon's function is related to the
NP-Weyl scalar through the relation,
\begin{equation}
  \label{eq:PdeLTOW}
  \Psi_{2}^{\np} = -\f{W}{r^{3}}. 
\end{equation}
This relation \(\Psi_{2}^{\np}\) will not be plotted
and instead the reader is referred  to Figure~\ref{fig:PdLW} for \(W(r).\)
The vanishing of most of the Weyl scalars~\eqref{eq:psi013} and
\(\Psi_{2}\)
being non-zero also proves that Tolman VII is a solution of Petrov
type D: It has two repeated principal null vectors: \(\vec{l}_{a},\)
and \(\vec{n}_{a},\)
expected from a static and spherically symmetric solution.

\begin{figure}[h]
   \centering
   \begin{subfigure}[b]{0.5\textwidth}
       \includegraphics[width=\textwidth]{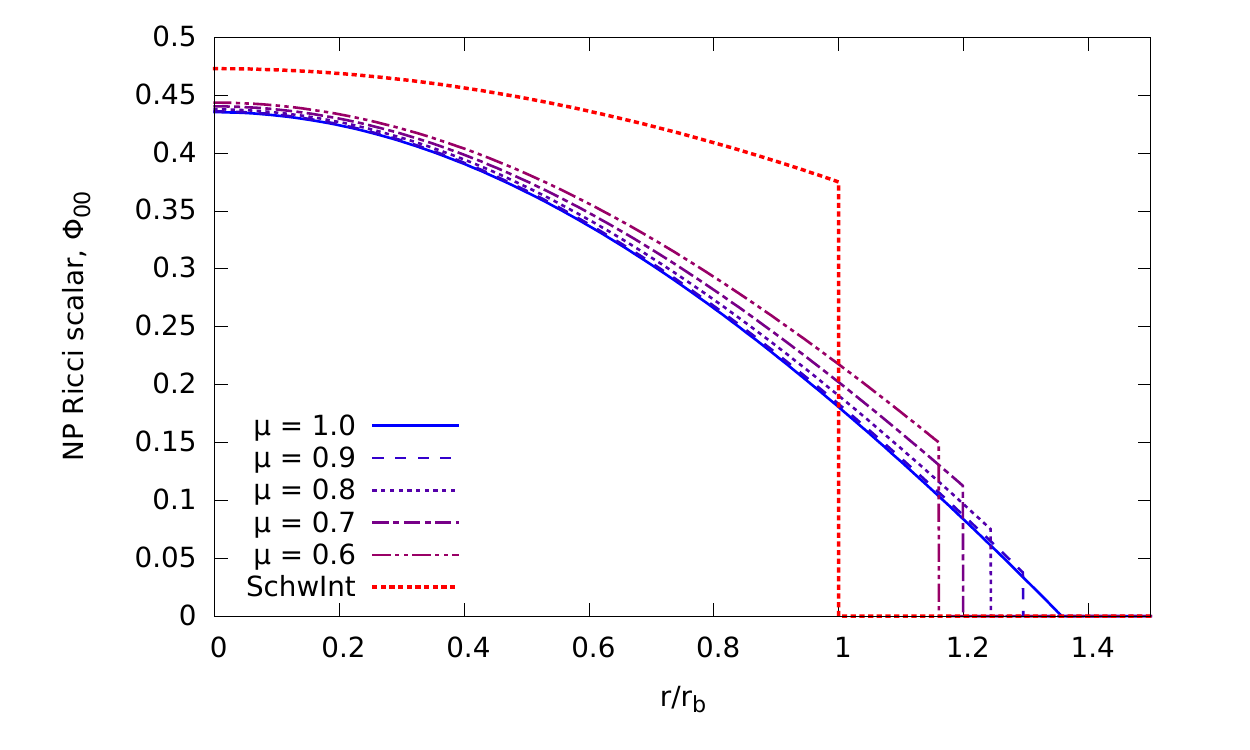}
       \caption{Constant central density}\label{fig:NPPhi00R}
   \end{subfigure}~
   \begin{subfigure}[b]{0.5\textwidth}
       \includegraphics[width=\textwidth]{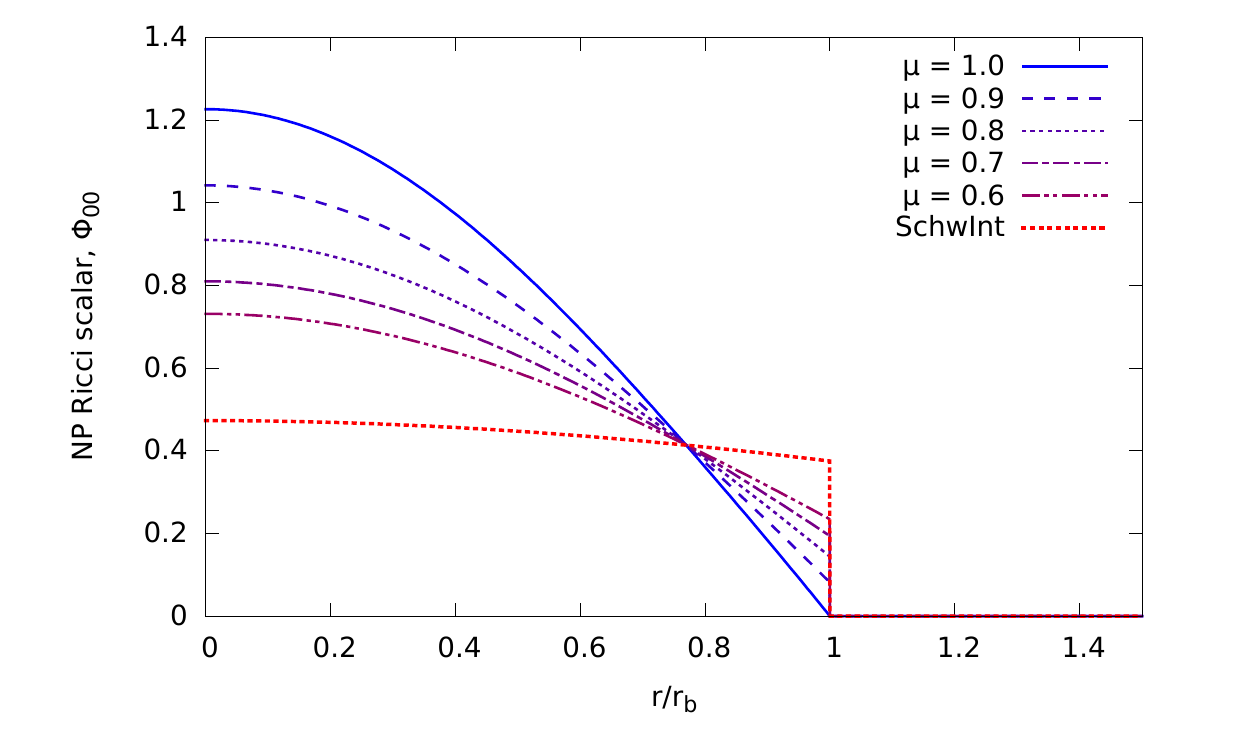}
       \caption{Constant radius}\label{fig:NPPhi00M}
   \end{subfigure}
   \caption{(Colour online) The $\Phi_{00}^{\np}$ NP Ricci scalar with the
     radial coordinate. The parameter values are
     $M = 1/4, \rho_{c} = 3/(16\pi)$ and $\mu$ taking the various
     values shown in the legend for figure~(\subref{fig:PdLWR}), and
     $M = 1/4$ for figure~(\subref{fig:PdLWM}).  Also shown
     for comparison is the Schwarzschild interior metric for the same
     parameter values.}
\label{fig:Phi00}
\end{figure}

\begin{figure}[h]
   \centering
   \begin{subfigure}[b]{0.5\textwidth}
       \includegraphics[width=\textwidth]{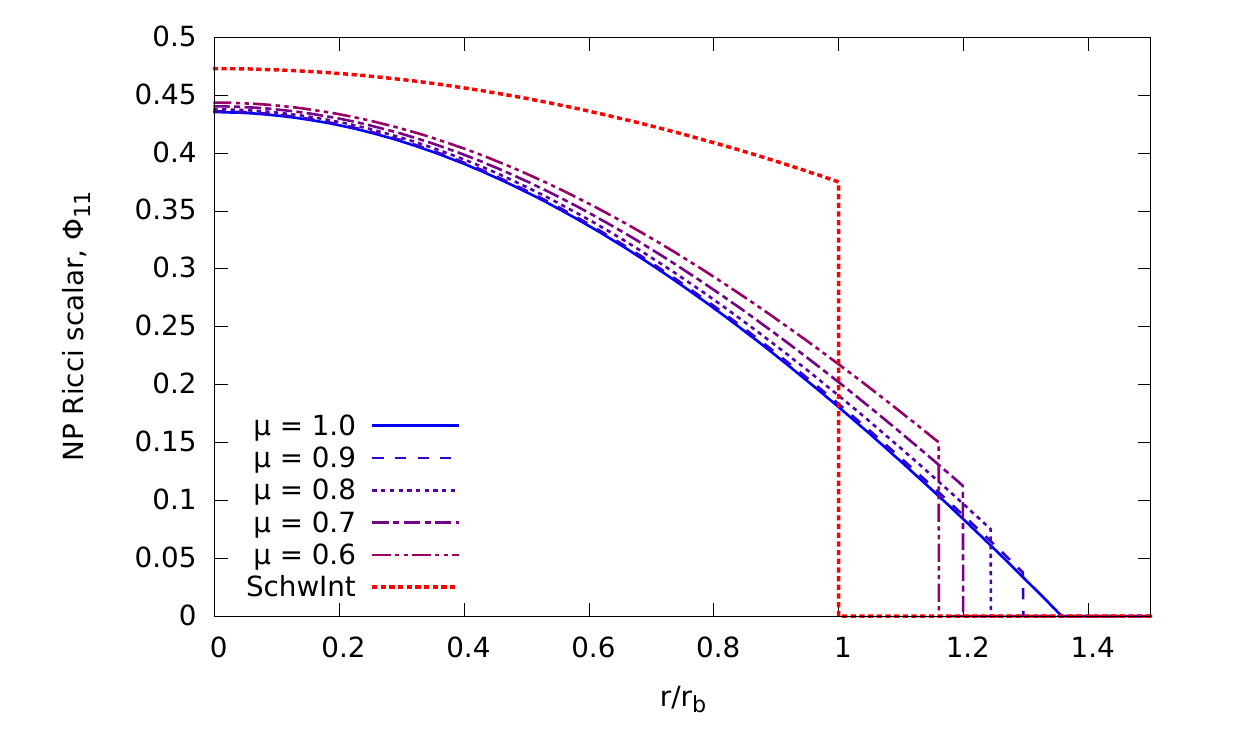}
       \caption{Constant central density}\label{fig:NPPhi11R}
   \end{subfigure}~
   \begin{subfigure}[b]{0.5\textwidth}
       \includegraphics[width=\textwidth]{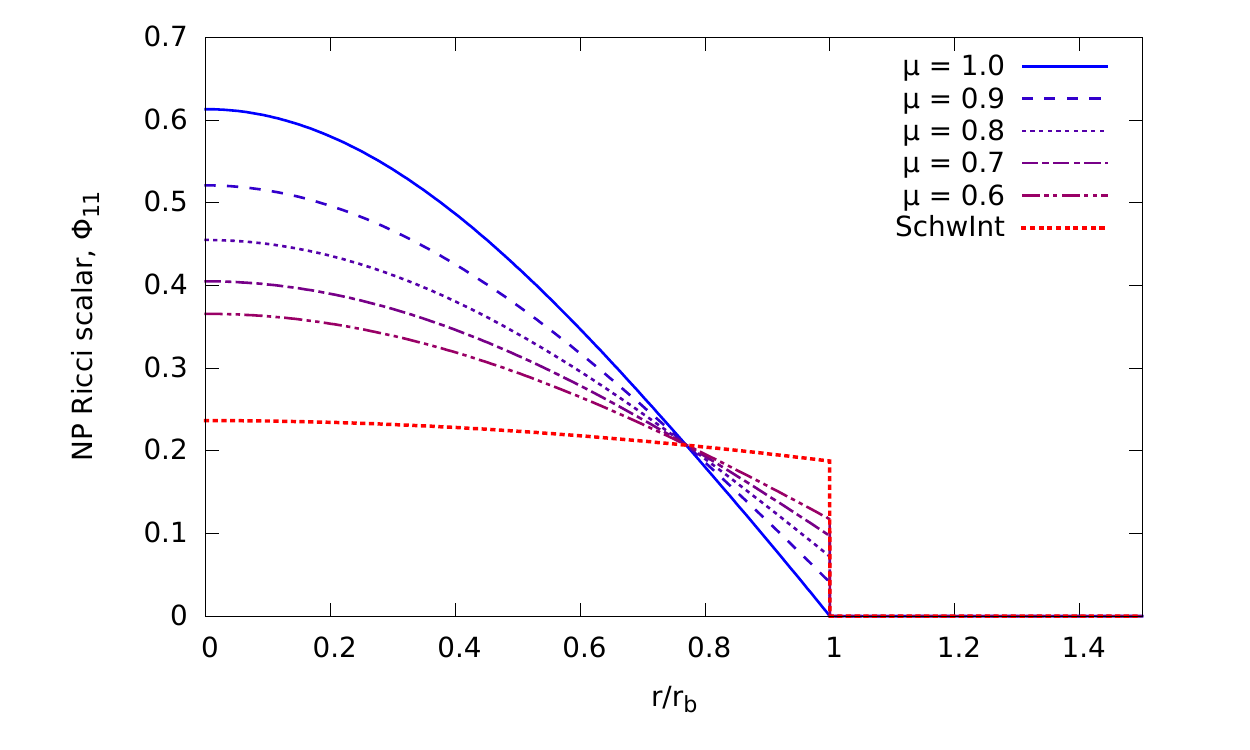}
       \caption{Constant radius}\label{fig:NPPhi11M}
   \end{subfigure}
   \caption{(Colour online) The $\Phi_{11}^{\np}$ NP Ricci scalar with the
     radial coordinate. The parameter values are
     $M = 1/4, \rho_{c} = 3/(16\pi)$ and $\mu$ taking the various
     values shown in the legend for figure~(\subref{fig:PdLWR}), and
     $M = 1/4$ for figure~(\subref{fig:PdLWM}). Also shown
     for comparison is the Schwarzschild interior metric for the same
     parameter values.}
\label{fig:Phi11}
\end{figure}

Continuing to look at the other scalar functions of the NP-formalism
by turning to the Ricci scalar functions, one finds
that some of them vanish identically:
\begin{equation}
\Phi_{01}^{\np} = \Phi_{02}^{\np} = \Phi_{12}^{\np} = 0,  
\end{equation}
while the non-vanishing ones are given by
\begin{equation}
\Phi_{00}^{\np} = \Phi_{22}^{\np} = \f{b}{2} - ar^{2} + \f{ar^{4} - br^{2} + 1}{2r} \left\{ \deriv{}{r} 
  \left[ \log{(c_{1} \cos{(\phi \xi)} + c_{2} \sin{(\phi \xi)} )} \right] \right\},  
\end{equation}
and these are shown in Figure~\ref{fig:Phi00} and,
\begin{multline}
\Phi_{11}^{\np} =\f{1}{4} \left\{ b - a r^{2} + 
  \left[ \f{1-br^{2}+ar^{4}}{c_{1} \cos{(\phi \xi)} + c_{2} \sin{(\phi \xi)} }\right]  
  \sderiv{}{r} (c_{1} \cos{(\phi \xi)} + c_{2} \sin{(\phi \xi)} ) +\right. \\ 
  + \left. \left( 2ar^{3} - br \right) \deriv{}{r} 
  \left[ \log{(c_{1} \cos{(\phi \xi)} + c_{2} \sin{(\phi \xi)} )} \right] \right\},  
\label{eq:Phi11}
\end{multline}
which is shown in Figure~\ref{fig:Phi11}.  The two graphs, and
functions are related to each other by
\begin{equation}
\Phi_{11} = \f{\Phi_{00}}{2},
\label{eq:PhiRel}  
\end{equation}
and this is a result that can be shown through the isotropy relation
\(G_{r}{}^{r} = G_{\theta}{}^{\theta}.\)
Using the latter, and substituting for the second derivative
in~\eqref{eq:Phi11} results in~\eqref{eq:PhiRel} 
The NP Ricci scalar which is related to the Ricci scalar is given by
\begin{multline}
  \Lambda = \f{1}{12} \left\{3b - 5ar^{2} 
    -\left[ \f{1-br^{2}+ar^{4}}{c_{1} \cos{(\phi \xi)} + c_{2} \sin{(\phi \xi)}}\right] \sderiv{}{r} [c_{1} \cos{(\phi \xi)} + c_{2} \sin{(\phi \xi)}] + \right.\\
\left. + \left[\f{3br^{2} - 4ar^{4} -2}{12r} \right] \deriv{}{r} \left[ \log{(c_{1} \cos{(\phi \xi)} + c_{2} \sin{(\phi \xi)} )} \right]
\right\},
\end{multline}
and is shown in Figure~\ref{fig:NPLambda}.
\begin{figure}[h]
   \centering
   \begin{subfigure}[b]{0.5\textwidth}
       \includegraphics[width=\textwidth]{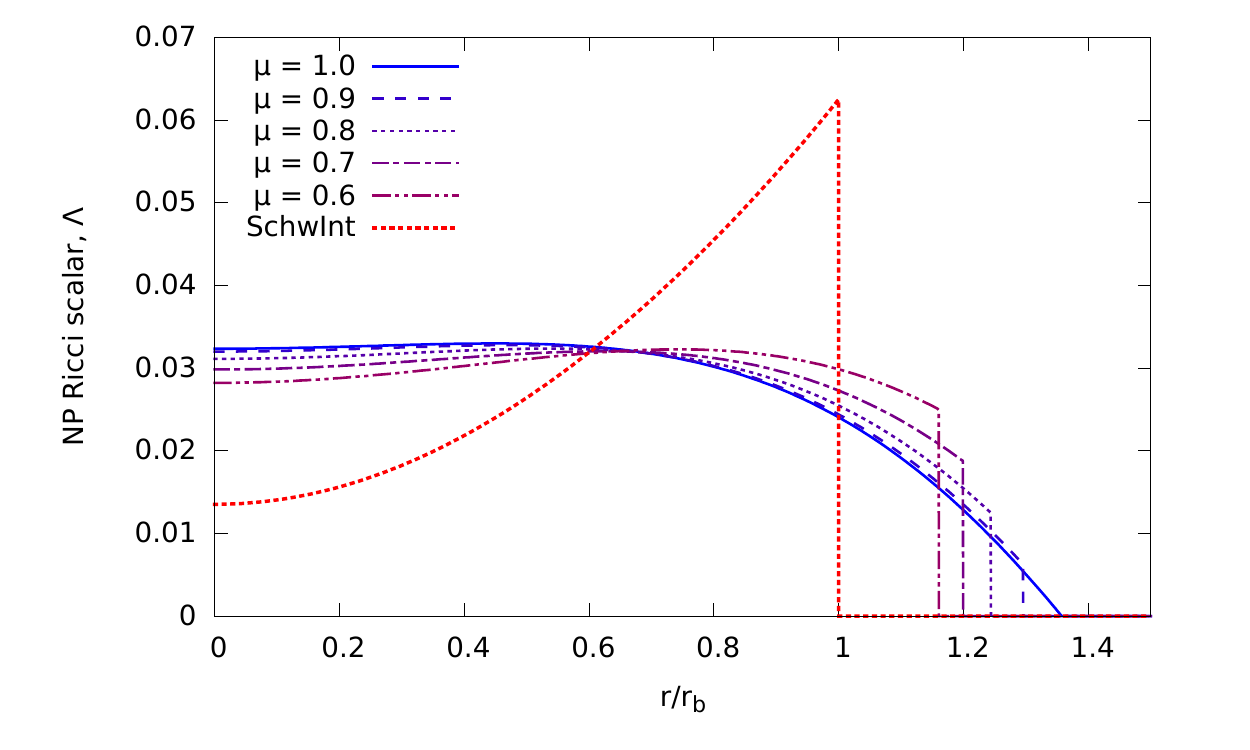}
       \caption{Constant central density}\label{fig:NPLambdaR}
   \end{subfigure}~
   \begin{subfigure}[b]{0.5\textwidth}
       \includegraphics[width=\textwidth]{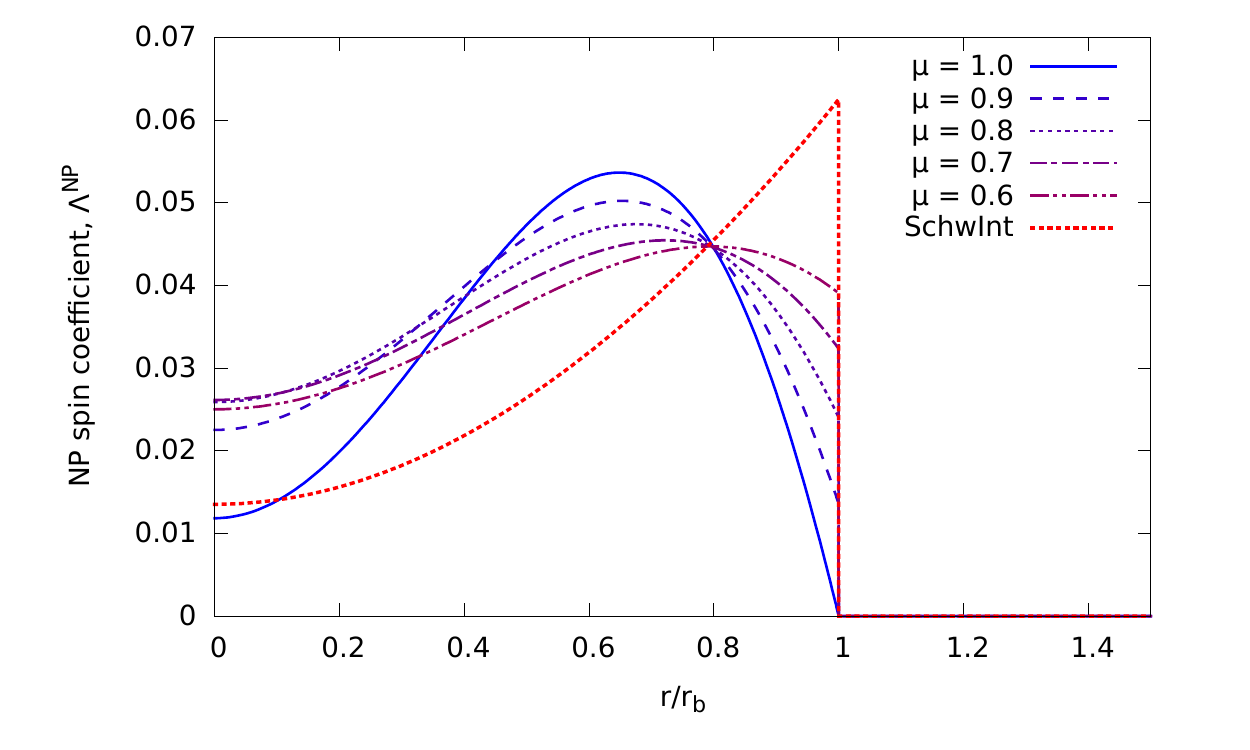}
       \caption{Constant radius}\label{fig:NPLambda}
   \end{subfigure}
   \caption{(Colour online) The $\Lambda^{\np}$ NP Ricci scalar with the
     radial coordinate. The parameter values are
     $M = 1/4, \rho_{c} = 3/(16\pi)$ and $\mu$ taking the various
     values shown in the legend for figure~(\subref{fig:PdLWR}), and
     $M = 1/4$ for figure~(\subref{fig:PdLWM}).  Also shown
     for comparison is the Schwarzschild interior metric for the same
     parameter values.}
\label{fig:NPLambda}
\end{figure}
This completes the list of NP scalars and rotation coefficients.  
% The
% relation~\eqref{eq:PdeLTOW} should come to no surprise since the
% behaviour of \(\Psi_{2}^{\np}\)
% is that~\cite{NewPen62}, \[\Psi_{2} \sim \mathcal{O}(r^{-3}),\]
% asymptotically, so that \(W\)
% is just the proportionality factor in the latter equation.
\clearpage
\bibliography{bibliography1}
\end{document}